\documentclass[%
reprint,
amsmath,amssymb,
aps,
pra,
]{revtex4-2}

\usepackage{graphicx}
\usepackage{dcolumn}
\usepackage{bm}

\begin{document}


\title{Quantum Commutation Relationship for Photonic Orbital Angular Momentum}

\author{Shinichi Saito}
 \email{shinichi.saito.qt@hitachi.com}
\affiliation{Center for Exploratory Research Laboratory, Research \& Development Group, Hitachi, Ltd. Tokyo 185-8601, Japan.}

\date{\today}

\begin{abstract}
Orbital Angular Momentum (OAM) of photons are already ubiquitously being used for numerous applications. 
However, there is a fundamental question whether photonic OAM operators satisfy standard quantum mechanical commutation relationship or not; this also poses a serious concern on the interpretation of an optical vortex as a fundamental quantum degree of freedom.
Here, we examined canonical angular momentum operators defined in a cylindrical coordinate,  and applied them to Laguerre-Gauss (LG) modes in a graded index (GRIN) fibre.
We confirmed the validity of commutation relationship for the LG modes and found that ladder operators also work properly with the increment or decrement in units of the Dirac constant ($\hbar$).
With those operators, we calculated the quantum-mechanical expectation value of the magnitude of angular momentum, which includes contributions from both intrinsic and extrinsic OAM.
The obtained results suggest that OAM characterised by the LG modes exhibits a well-defined quantum degree of freedom.
\end{abstract}

\maketitle


\section{Introduction}
Quantum commutation relationship between operators is an indispensable characteristic in connection with measurements of physical observables \cite{Dirac30, Baym69,Sakurai14,Sakurai67}.  Angular momentum operators are especially important as generators of rotation for states with angular momentum state, $|m \rangle$, which is characterised with a quantised integer or a half-integer, $m$, along a certain direction (say, $z$) \cite{Dirac30, Baym69,Sakurai14,Sakurai67}.
The states pointing directions such as $x$- and $y$-directions or any other directions in three-dimensional ($3D$) space are described by superposition states of orthogonal basis states \cite{Dirac30, Baym69,Sakurai14,Sakurai67}, and commutation relationship is used to rotate the states.
The spin, $\hat{\bf S}$, with one-half in units of the Dirac constant, $\hbar=h/(2 \pi)$, where $h$ it the Plank constant, for an electron is perfectly described with Pauli's spin matrices \cite{Dirac30,Baym69,Sakurai14,Sakurai67}. 
The Orbital Angular Momentum (OAM), $\hat{\bf L}$, for an electron trapped in spherical potential, e.g. an electron in an atom, is also characterised with an integer quantum number in units of $\hbar$, where an orbital, such as $s$, $p$, $d \cdots$, described by spherical harmonics, $Y_{l}^{m}$, satisfies $\hat{\bf L}^2|l, m \rangle=l(l+1)\hbar^2 |l, m \rangle$, with an integer $l$ associated with the magnitude of angular momentum \cite{Dirac30, Baym69,Sakurai14,Sakurai67}.
Furthermore, the total angular momentum, $\hat{\bf J}=\hat{\bf S}+\hat{\bf L}$, of their simple sum is also known as a well-defined quantum angular momentum operator.
In general, those angular momentum operators are described by elegant mathematics - Lie algebra, which is looked upon as a triumph in the mathematical formulation of quantum mechanics \cite{Dirac30, Baym69,Sakurai14,Sakurai67}.

However, this elegant theoretical framework of angular momentum is less trivial in applying to a photon \cite{Allen92,Enk94,Leader14,Barnett16,Yariv97,Jackson99,Grynberg10,Bliokh15}, since a photon is usually uni-directionally propagating (say, along $z$).
Therefore, the apparent spherical rotational symmetry is absent for a photon, such that the situation is rather different from that of an electron confined in a spherically symmetric $3D$ space.
The propagation direction of a photon sets a natural quantisation axis; in the propagation direction, the angular momentum operator, $\hat{L}_z$, is successfully obtained by the classical analogue using the Poynting vector \cite{Allen92,Enk94,Allen00,Leader14,Barnett16,Grynberg10,Bliokh15}.
However, angular momentum operators along the directions perpendicular to the propagation direction ($x$-$y$ plane) are not defined uniquely in a gauge-invariant way \cite{Enk94,Leader14,Barnett16}.
It was argued that the spin and OAM operators would commute \cite{Enk94}.
It is now generally believed that the total angular momentum operator $\hat{\bf J}$ for photons is well-defined, while it is impossible to split into $\hat{\bf S}$ and $\hat{\bf L}$ in a local gauge-invariant way \cite{Leader14,Barnett16,Ji10}, despite a big challenge against this claim \cite{Chen08}. This is outside the scope of this paper, and we will revisit this issue in a forthcoming paper.

Nevertheless, there are several naive questions, which should be addressed:
(1) The spin of a photon could be proper angular momentum, which would satisfy commutation relationship, at least in the absence of OAM.
The selection rules of absorption and excitation of a photon in materials \cite{Baym69,Sakurai14,Sakurai67,Yariv97,Chuang09} are clear evidence to expect that the spin of a photon is transferred to angular momenta of an electron.
If the spin of a photon is conserved, while accepting well-defined OAM of an electron, why could the spin of a photon be regarded as a classical degree of freedom, which is described by a commutable operator?  
It should be treated with appropriate quantum commutation relationship, if spin of a photon is a proper quantum mechanical degree of freedom.
Moreover, what about the relationship between spin of a photon and the polarisation \cite{Goldstein11,Gil16}?  
Detailed discussion about this will be provided in a separate paper.
(2) A photon with OAM carries quantised angular momentum of $\hbar m$ along the direction of propagation, which was successfully described by a Laguerre-Gauss (LG) mode in a cylindrical coordinate \cite{Allen92}.  
Here, some questions arise: 
Why $\hbar$ appeared in OAM, which is usually the evidence of quantisation? 
If the OAM is a classical degree of freedom, described by a commutable operator, we normally expect that $\hbar$ would not appear, which is clearly not the case.
Can we define standard quantum mechanical canonical angular momentum operators and apply them to LG modes?
Moreover, what happens if we define ladder operators for raising or lowering angular momentum in a standard way, such as $\hat{L}_{\pm}=\hat{L}_x \pm i \hat{L}_y$, and apply them to the LG modes?
Can we show the increment or decrement of quantised angular momentum in units of $\hbar$?  These are non-trivial questions.  
In this paper, we will answer to those questions on OAM by directly calculating the matrix elements.

Here is the outline of this paper: 
In Sec.~II, we review fundamental principles and equations, which we are relying on, and explain our model.
We are interested in photonics that can be applied to communication technologies and low-energy condensed-matter physics.
Consequently, we will not deal with high-energy physics nor those issues related to Lorentz invariance \cite{Leader14,Barnett16,Chen08,Ji10} in this paper.
We are considering monochromatic coherent light sources from lasers, such that the incoherent unpolarised lights will not be considered, either.
We have also included detailed appendices to define associate Laguerre functions and have shown various mathematical formulas to make this paper self-contained.
In the calculations, well-known formulas and equations appear; yet the others are newly derived, particularly in evaluating the ladder operations.
We hope the appendices will help readers, because definitions depend on literatures for factors and signs.

In Sec.~III, we explain our methods to evaluate OAM operators. 
We clarify the challenges to apply OAM operators to plane-waves, which are not successful.
Nevertheless, this would help to understand the problem, which we would like to address.
There, we show that the main problem of the plane-waves is a lack of a node at the core of the waveguide, which is also called as topological charge.
We show that this problem is solved by using LG modes. 
In Sec.~IV, we show our main calculation results of various matrix elements for OAM and discuss their implications.
Our results show that the LG modes actually satisfy the quantum canonical commutation relationship of angular momentum.
We also obtained expectation values of the magnitude of OAM.
Finally, in Sec.~V, we conclude that OAM is indeed a genuine quantum-mechanical observable at least in a graded index (GRIN) fibre \cite{Kawakami68,Yariv97} satisfying some conditions.

\section{Principles and Models}
\subsection{Maxwell's equations}
We start from Maxwell's equations \cite{Jackson99,Yariv97},
\begin{eqnarray}
\nabla \times \bm{\mathcal E} &=-\frac{\partial \bm{\mathcal B}}{\partial t} \\
\nabla \times \bm{\mathcal H} &=\frac{\partial \bm{\mathcal D}}{\partial t} \\
\nabla \cdot \bm{\mathcal D} &=0 \\
\nabla \cdot \bm{\mathcal B} &=0,
\end{eqnarray}
in a non-magnetic transparent material of the dielectric constant $\epsilon$ and the permittivity $\mu_0$ without charges and currents.
As usual for describing electromagnetic fields \cite{Jackson99,Yariv97}, we use complex oscillation fields in SI units for electric field $\bm{\mathcal E}$ (V/m), displacement $\bm{\mathcal D}$ (C/m$^2$), magnetic field $\bm{\mathcal H}$ (A/m), and induction $\bm{\mathcal B}$ (Wb/m$^2$), with materials equations $\bm{\mathcal D}=\epsilon \bm{\mathcal E}$ and $\bm{\mathcal B}=\mu_0 \bm{\mathcal H}$.
All physical observables must be expected to be real without the imaginary part \cite{Sakurai14}  such that experimentally observable fields should be considered by taking the real part, such as for electric field $\bm{E}=(\bm{\mathcal E}+\bm{\mathcal E}^*)/2$, after the calculation using a complex field of $\bm{\mathcal E}$.
While $\mu_0$ is approximately the same as that in a vacuum, $\epsilon$ is different in a material from the value of $\epsilon_0$ in a vacuum \cite{Jackson99,Yariv97}.
In a non-uniform material, $\epsilon=\epsilon({\bf r})$ depends on a position ${\bf r}=(x,y,z)$.
If we assume that the profile of $\epsilon$ is sufficiently uniform ($\nabla \epsilon \simeq  0$) compared with the size of wavelength, $\lambda$, of a photon, we obtain Helmholtz equation, 
\begin{eqnarray}
\nabla^2 \bm{\mathcal E}
=
\mu_0 \epsilon 
\frac{\partial^2}{\partial t^2}{\bm{\mathcal E}},
\end{eqnarray}
which is valid in a perfectly uniform material and in a vacuum.
In this paper, we will examine the Helmholtz equation in more detail as follows, but we will not examine its validity any further.
The only source of approximations are the sufficient uniformity ($\nabla \epsilon \simeq  0$). 
Therefore our analysis will not be valid if $\epsilon$ is significantly changed in nano-metre-scale such as for photonic crystals \cite{Joannopoulos08,Sotto18,Sotto18b,Sotto19} and other inhomogeneous systems \cite{Bliokh17,Bliokh17b}.

\subsection{Mapping to Schr\"odinger equation}
First, we see the qualitative feature of the Helmholtz equation in a uniform material by assuming a solution for a linearly polarised monochromatic plane wave
\begin{eqnarray}
\bm{\mathcal E}(x,y,z)
=
E_0
\psi (x,y,z)
{\rm e}^{i(kz-\omega t)} \hat{\bf n},
\end{eqnarray}
where $E_0$ is the magnitude of the electric field, $k$ is the wavenumber in the material, $\omega$ is the angular frequency, the unit vector $\hat{\bf n}$ is the direction of the polarisation in the $(x,y)$ plane due to the transversality of the electromagnetic wave, and $\psi (x,y,z)$ is the envelope wavefunction of a photon.
Inserting $\bm{\mathcal E}$ into the Helmholtz equation, we obtain
\begin{eqnarray}
\left(
\frac{\partial^2 }{\partial x^2} 
+
\frac{\partial^2 }{\partial y^2} 
+
\frac{\partial^2 }{\partial z^2}
+2ik
\frac{\partial }{\partial z}
-k^2
\right)
\psi
=
-\mu_0
\epsilon
\omega^2
\psi.
\end{eqnarray}
For various practical applications in laser optics, rays from laser sources are sufficiently collimated, such that the rays can be regarded as paraxial beams \cite{Yariv97}.
In such a case, we can use slowly varying approximation \cite{Yariv97} to neglect the second derivative,
\begin{eqnarray}
\frac{\partial^2 \psi}{\partial z^2}
\ll
k \frac{\partial \psi}{\partial z}
, \ 
k^2 \psi , 
\end{eqnarray}
and obtain \cite{Barnett16b}
\begin{eqnarray}
i
\lambdabar
\frac{\partial }{\partial z}
\psi
=
-
\frac{\lambdabar^2}{2n}
\left(
\frac{\partial^2 }{\partial x^2} 
+
\frac{\partial^2 }{\partial y^2} 
\right)
\psi, \label{SVA}
\end{eqnarray}
where we have used the dispersion relationship, $\omega =v k$, with the velocity $v=1/\sqrt{\mu_0 \epsilon}=c/n$ of a photon in a material of a refractive index of $n$.
The velocity of a photon in a vacuum is $c=1/\sqrt{\mu_0 \epsilon_0}$ and $\lambdabar=\lambda/(2\pi)$ is an angular wavelength.
Equation (\ref{SVA}) is exactly the same form with a standard non-relativistic Schr\"odinger equation \cite{Simon93,Barnett16b,Baym69,Sakurai14,Chuang09}
\begin{eqnarray}
i
\hbar
\frac{\partial }{\partial t}
\psi
=
-
\frac{\hbar^2}{2m}
\left(
\frac{\partial^2 }{\partial x^2} 
+
\frac{\partial^2 }{\partial y^2} 
\right)
\psi
\end{eqnarray}
for a particle of mass, $m$, in a $2D$ $xy$-plane at time $t$.
The correspondence is summarised in Table \ref{tab:table1}.

\begin{table}[h]
\caption{\label{tab:table1}
Mapping of Helmholtz equation to a non-relativistic Schr\"odinger equation.
}
\begin{ruledtabular}
\begin{tabular}{cc}
\textrm{Helmholtz Eq.} & \textrm{Schr\"odinger Eq.}\\
\colrule
\textrm{Space:} $z$& \textrm{Time:} $t$ \\
\textrm{Refractive index:} $n$ & \textrm{Mass:} $m$\\
\textrm{Angular wavelength:} $\lambdabar$  & \textrm{Dirac constant:} $\hbar$\\
\end{tabular}
\end{ruledtabular}
\end{table}

This implies the propagation of a photon along a paraxial optical path can be described by the same mechanism with the dynamics of a {\it massive} quantum-mechanical particle \cite{Barnett16b}.
In fact, electron vortices similar to photonic ones were observed \cite{Lloyd17}, and essentially the same mathematical and physical techniques are applicable to both electronic and photonic systems.
It is also intuitive to recognise $n$ for a photon corresponds to $m$ of the particle, such that $v$ is low in a material of large $n$ similar to low velocity of a heavy particle.

\subsection{Coherent state for photons}
On the contrary to the above similarity between electrons and photons, the important difference is coming from the nature of statistics between Fermions and Bosons.
For photons, we are considering monochromatic rays from lasers, which are considered to be in a macroscopic coherent state, exhibiting Bose-Einstein condensation, enabled by the Bose statistics for spin integer particles \cite{Dirac30, Sakurai67,Grynberg10,Fox06}.
On the other hand, electrons are Fermions due to their spin $1/2$ characteristics \cite{Dirac30, Sakurai67}, such that a macroscopic coherence is not expected, except for ordered states, such as a superconducting state, which is similar to Bose-Einstein condensation of Cooper pairs \cite{Schrieffer71}.
Here, we consider a coherent state for photons to understand the quantum-mechanical state with certain polarisation and orbital angular momentum.
A coherent state cannot be described by a fixed number state only due to the phase coherence.
Instead, a coherent state is described by a fixed phase, while allowing fluctuation in the number of photons from their average value by a superposition of states with different number of states \cite{Grynberg10,Fox06}.
Specifically, a coherent state for $\sigma$-polarisation in a uniform material is described by
\begin{eqnarray}
|\alpha_{\sigma} \rangle
&=&{\rm e}^{-\frac{|\alpha_{\sigma}|^2}{2}}
{\rm e}^{\alpha_{\sigma} \hat{a}_{\sigma}^{\dagger}}
|0\rangle \nonumber \\
&=&{\rm e}^{-\frac{|\alpha_{\sigma}|^2}{2}}
\sum_{n_{\sigma}=0}^{\infty}
\frac{(\alpha_{\sigma} \hat{a}_{\sigma}^{\dagger})^{n_{\sigma}}}{n_{\sigma}!}
|n_{\sigma}\rangle,
\end{eqnarray}
where $\sigma={\rm H}$ for horizontally polarised state and  $\sigma={\rm V}$ for vertically polarised state. $\alpha_{\sigma}$ is a complex number.
We use horizontally or vertically polarised states as basis states, for simplicity, but in general we can take other orthonormal bases such as left/right polarised states and diagonal/anti-diagonal states \cite{Yariv97}. 
$\hat{a}_{\sigma}^{\dagger}$ and $\hat{a}_{\sigma}$ are creation and annihilation operators, satisfying Bose commutation relationship \cite{Sakurai67,Grynberg10,Fox06}
\begin{eqnarray}
[\hat{a}_{\sigma},\hat{a}_{\sigma^{'}}^{\dagger}]=\delta_{{\sigma},{\sigma}^{'}},
\end{eqnarray}
where $\delta$ is the Kronecker delta.
A coherent state is best characterised by the fact that it is an eigenstate of an annihilation operator, which can be directly confirmed by the commutation relationship as 
 \begin{eqnarray}
\hat{a}_{\sigma}|\alpha_{\sigma} \rangle
&=&{\rm e}^{-\frac{|\alpha_{\sigma}|^2}{2}}
\sum_{n_{\sigma}=1}^{\infty}
\frac{\alpha^{n_{\sigma}}}{\sqrt{n_{\sigma}!}}
\sqrt{n_{\sigma}}|n_{\sigma}-1\rangle \nonumber \\
&=&\alpha_{\sigma}{\rm e}^{-\frac{|\alpha_{\sigma}|^2}{2}}
\sum_{n_{\sigma}=0}^{\infty}
\frac{\alpha^{n_{\sigma}}}{\sqrt{n_{\sigma}!}}
|n_{\sigma} \rangle \nonumber \\
&=&\alpha_{\sigma}|\alpha_{\sigma} \rangle.
\end{eqnarray}
We can also confirm that $|\alpha_{\sigma} \rangle$ is normalised as $\langle \alpha_{\sigma} |\alpha_{\sigma} \rangle =1$.
Then, we can calculate the average number of photons for both polarised states as 
\begin{eqnarray}
\langle \hat{a}_{\rm H}^{\dagger} \hat{a}_{\rm H} \rangle
&=|\alpha_{\rm H}|^2=N_{\rm H}=N\cos^2 \alpha\\
\langle \hat{a}_{\rm V}^{\dagger} \hat{a}_{\rm V} \rangle
&=|\alpha_{\rm V}|^2=N_{\rm V}=N\sin^2 \alpha,
\end{eqnarray}
where $N_{\rm H}$ and $N_{\rm V}$ are the average number of photons for horizontally and vertically polarisation, respectively, $N=N_{\rm H}+N_{\rm V}$ is the total number of photons, and $\alpha$ is the auxiliary angle ($0\le \alpha \le \frac{\pi}{2}$) to describe the polarisation.
Alternatively, $\alpha_{\sigma}$ is determined by the polarisation state as
\begin{eqnarray}
\alpha_{\rm H}
&=&\sqrt{N} \cos \alpha \\
\alpha_{\rm V}
&=&\sqrt{N} \sin \alpha \ 
{\rm e}^{i \delta},
\end{eqnarray}
where $\delta \in (0,2\pi)$ is the phase of the polarisation.
The overall coherent state is described by the direct product as 
\begin{eqnarray}
|\alpha_{\rm H},\alpha_{\rm V}\rangle
=|\alpha_{\rm H}\rangle | \alpha_{\rm V}\rangle.
\end{eqnarray}
 
Now, we have prepared the coherent state, and the next step is to consider the quantum many-body description of the electromagnetic field, which is achieved by considering the following complex electric field operator
\begin{eqnarray}
\bm{\hat{\mathcal{E}}}(z,t)=
\sqrt{
  \frac{2 \hbar \omega}{\epsilon V}
  }
{\rm e}^{i \beta}
\left(
  \hat{a}_{\rm H}
  \hat{\bf x}
  +\hat{a}_{\rm V}
  \hat{\bf y}
\right),
 \end{eqnarray}
where $V$ is the volume, $ \hat{\bf x}$ and $ \hat{\bf y}$ are unit vectors along $x$ and $y$, and $\beta=kz-\omega t + \beta_0$ describes the trivial time and space evolution with a global $U(1)$ phase of $\beta_0$.
If we apply $\hat{\mathcal{E}}(z,t)$ to the coherent state $|\alpha_{\rm H},\alpha_{\rm V}\rangle$ as, 
\begin{eqnarray}
\bm{\hat{\mathcal{E}}}(z,t)
|\alpha_{\rm H},\alpha_{\rm V}\rangle
&=&
\bm{\mathcal{E}}(z,t)
|\alpha_{\rm H},\alpha_{\rm V}\rangle, 
 \end{eqnarray}
we realise the state is an eigenstate of $\hat{\mathcal{E}}(z,t)$ with the complex eigenvalue of 
\begin{eqnarray}
\bm{\mathcal{E}}(z,t)
&=
E_{0}
{\rm e}^{i \beta}
\left(
  \cos \alpha
  \hat{\bf x}
  +
  {\rm e}^{i \delta}
  \sin \alpha
  \hat{\bf y}
\right),
 \end{eqnarray}
where $E_0=\sqrt{2 \hbar \omega N/(\epsilon V)}$.
This means that the coherent state is a simultaneous eigenstate for diagonalising $\bm{\hat{\mathcal{E}}}(z,t)$ and $\hat{a}_{\sigma}$.
Alternatively, we can consider the coherent state, $\bm{\hat{\mathcal{E}}}(z,t)|\alpha_{\rm H},\alpha_{\rm V}\rangle$, describes the photonic state of the system, since the multiplication of the operator does not change the state.
In fact, $\bm{\mathcal{E}}(z,t)$ actually corresponds to the spinor description of the polarisation state as 
\begin{eqnarray}
\left (
  \begin{array}{c}
    \mathcal{E}_{x} \\
    \mathcal{E}_{y}
  \end{array}
\right)
&=
E_{0}{\rm e}^{i\beta}
\left (
  \begin{array}{c}
    \cos \alpha \ \ \ \  \\
    \sin \alpha \ {\rm e}^{i\delta}
  \end{array}
\right),
 \end{eqnarray}
where the matrix part is nothing but a Jones vector \cite{Yariv97} to describe the polarisation state of a photon.
Moreover, the overall factor of the complex electric field of $E_{0}{\rm e}^{i\beta}$ corresponds to the orbital part of the wavefunction for photons in a uniform material, which is the solution of Helmholtz equation.

More generally, for describing a coherent monochromatic ray from a laser source propagating in a waveguide or a fibre, the complex electric field operator must be defined as  
\begin{eqnarray}
\bm{\hat{\mathcal{E}}}({\bf r},t)
&=&
\mathcal{E}({\bf r},t)
\left(
  \hat{a}_{\rm H}
  \hat{\bf x}
  +\hat{a}_{\rm V}
  \hat{\bf y}
\right),
\end{eqnarray}
where the {\it scalar} complex electric field, $\mathcal{E}({\bf r},t)=E_0\psi({\bf r}) {\rm e}^{i\beta}$, describes the orbital part. 
The state, $\bm{\hat{\mathcal{E}}}({\bf r},t)|\alpha_{\rm H},\alpha_{\rm V}\rangle$, describes the entire photonic state, including polarisation.
In this case, $\bm{\mathcal{E}}({\bf r},t)$ becomes
\begin{eqnarray}
\left (
  \begin{array}{c}
    \mathcal{E}_{x} \\
    \mathcal{E}_{y}
  \end{array}
\right)
&=
E_0 \Psi({\bf r}) {\rm e}^{i\beta}
\left (
  \begin{array}{c}
    \cos \alpha \ \ \ \  \\
    \sin \alpha \ {\rm e}^{i\delta}
  \end{array}
\right), 
 \end{eqnarray}
where the orbital part is described by $\Psi({\bf r})=\psi({\bf r}) {\rm e}^{i\beta}$.
$\Psi({\bf r})$ is determined by the {\it scalar} Helmholtz equation
\begin{eqnarray}
\nabla^2 
\Psi({\bf r})
=
\mu_0 \epsilon ({\bf r})
\frac{\partial^2}{\partial t^2}
\Psi({\bf r}),
\end{eqnarray}
and thus $\Psi({\bf r})$ is essentially a single-particle wavefunction, describing the orbital degree of freedom.
The reason why a macroscopic number of photonic state can be described by a single wavefunction comes from the Bose-Einstein condensed character of a superposition  state.
All photons are occupying a single states with fixed $\omega$, $k$, $\delta$, and $\alpha$, while allowing the fluctuation of the number of photons around its average value of $N$ using a coherent state. 
The polarisation state is also described as a superposition state of two orthogonal polarisation basis states, coming from intrinsic internal degrees of freedom described by a Jones vector.

The coherent state for a laser beam can also be written as 
\begin{eqnarray}
|N,\alpha, \delta \rangle
&=&
|\alpha_{\rm H},\alpha_{\rm V}\rangle \nonumber \\
&=&
{\rm e}^{-\frac{|\alpha_{\rm H}|^2}{2}}
{\rm e}^{\alpha_{\rm H} \hat{a}_{\rm H}^{\dagger}}
{\rm e}^{-\frac{|\alpha_{\rm V}|^2}{2}}
{\rm e}^{\alpha_{\rm V} \hat{a}_{\rm V}^{\dagger}}
|0\rangle \nonumber \\
&=&
{\rm e}^{-\frac{N}{2}}
{\rm e}^{\sqrt{N}(\cos \alpha \hat{a}_{\rm H}^{\dagger}
  +{\rm e}^{i \delta}\sin \alpha \hat{a}_{\rm V}^{\dagger})}
|0\rangle. 
\end{eqnarray}
If we want to calculate the real electric field, instead of the complex field, we should use the electric field operator defined by 
\begin{eqnarray}
\bm{\hat{
{\bf E}
}}
=
\frac{1}{2}
\left (
\bm{\hat{\mathcal{E}}}
+
\bm{\hat{\mathcal{E}}}^{\dagger}
\right),
\end{eqnarray}
which is an observable, such that we can calculate the expectation value, $\langle \bm{\hat{{\bf E}}} \rangle$, quantum-mechanically, using $|N,\alpha, \delta \rangle$.

\subsection{Laguerre-Gauss mode in a uniform material}
Above formalism is based on Maxwell equations, quantum statistics, and superposition principle. 
Therefore, it is virtually an exact consequence that $\Psi({\bf r})$ represents the wavefunction of coherent photonic states and satisfies the Helmholtz equation at least in a uniform material.
The similarity of quantum-mechanical nature of $\Psi({\bf r})$ was intuitively suggested in many pioneering works \cite{Allen92,Enk94,Allen00,Barnett16}.
Now, it becomes clearer that the intuitive correlation is not merely a coincidence but firmly supported by a quantum many-body theory rather than classical Maxwell's equations alone, since we cannot derive a wavefunction from classical mechanics.
Our formalism contains a standard vacuum state of Quantum Electro-Dynamics (QED) theory \cite{Sakurai67} in the limit of $n \rightarrow 1$, where $\Psi({\bf r})$ will become a simple plane wave, $\Psi({\bf r}) \rightarrow {\rm e}^{i\beta}$.

However, the plane wave is not the only solution of the Helmholtz equation, since a solution of differential equation depends also on the symmetry and boundary condition of the system \cite{Yariv97,Allen92}.
This is especially true in condensed matter physics, because a material is usually patterned in a specific form with a certain symmetry. 
Here, we derive a LG mode solution in a uniform material in a cylindrical coordinate by using the slowly varying approximation \cite{Yariv97,Allen92}.
For completeness, we will describe its full detail in this subsection.

Our starting point is the Helmholtz equation in a cylindrical coordinate $(r,\phi)$
\begin{eqnarray}
i\frac{\partial }{\partial z}
\psi
=
-
\frac{1}{2k}
\left(
\partial_{r}^{2}
+
\frac{1}{r} 
\partial_{r}
+
\frac{1}{r^2} 
\partial_{\phi}^{2}
\right)
\psi,
\end{eqnarray}
where $r=\sqrt{x^2+y^2}$ is the radius and $\phi=\tan^{-1}(y/x)$ is the azimuthal angle.
We will solve this equation by using a trial wavefunction
\begin{eqnarray}
\psi(r,\phi,z)
&=
\left (
  \frac{r}{w(z)}
\right)^{m}
f
\left(
\left(
  \frac{r}{w(z)}
\right)^{2}
\right)
{\rm  e}^{iP(z)+ik\frac{r^2}{2 q(z)}+im\phi+i\theta(z)},
\nonumber \\
\end{eqnarray}
where $w(z)$ is the beam-waist size, $P(z)$ is the phase shift for a beam expansion, $q(z)$ is the complex spherical radius, and $\theta(z)$ is another phase shift for radial and azimuthal expansions.
Here, we tentatively assume $m \ge 0$ for simplicity, and yet we relax this condition for all integer values, including negative values, at the end of the calculation.
While the trial wavefunction is inserted into the Helmholtz equation, it is useful to note that $\partial_x (fg)=g(\partial_x f)+f (\partial_x g)
=\psi(\partial_x f)/f+\psi \partial_x g/g$ holds.
Then, we obtain
\begin{eqnarray}
&2ki
\left(
iP^{'}
+\frac{1}{q}
\right)
-
\frac{k^2r^2}{q^2}
\left(
1-q^{'}
\right)
-2k\theta^{'} \nonumber \\
&
+\frac{4}{w^2}
\left(
  \frac{r}{w}
\right)^2
\frac{f^{''}}{f}
+
\frac{4ki}{q}
\left(
  \frac{r}{w}
\right)^2
\frac{f^{'}}{f}
+
\frac{2(2m+1)}{w^2}
\frac{f^{'}}{f}
+
\frac{2}{w^2}
\frac{f^{'}}{f} \nonumber \\
&-
2ki
\frac{mw^{'}}{w}
-4ki
\frac{r^2w^{'}}{w^3}
\frac{f^{'}}{f}
+
2ki
\frac{m}{q}
=0.
\end{eqnarray}
The Gaussian mode solution is given by assuming
\begin{eqnarray}
\frac{\partial P}{\partial z}
&=&  \frac{i}{q} \\
\frac{\partial q}{\partial z}
&=&1,
\end{eqnarray}
which will give us $q(z)=z+q_0=z-iz_0$ and $~P(z)=i\ln \left( 1+z/q_0 \right)$ with the confocal parameter $z_0=k w_0^2/2=\pi n w_0^2/\lambda$ and the minimum waist of $w_0$.
We then obtain the Gaussian factor
\begin{eqnarray}
{\rm e}^{ik\frac{r^2}{2 q(z)}}
=
{\rm e}^{-\frac{r^2}{w^2}}
{\rm e}^{i\frac{kr^2}{2R}}
\end{eqnarray}
and the phase-shift factor
\begin{eqnarray}
{\rm e}^{iP(z)}
=
\frac{w_0}{w}
{\rm e}^{-i \eta(z)},
\end{eqnarray}
where the beam waist, $w(z)$, the beam radius, $R(z)$, and the phase, $\eta(z)$, are given by
\begin{eqnarray}
w(z)
&=&w_0
\sqrt{
1+
  \left(
    \frac{z}{z_0}
   \right)^2
} \\
R(z)
&=&
z+    \frac{z_0^2}{z} \\
\eta(z)&=&\tan^{-1} (\frac{z}{z_0}).
\end{eqnarray}
The focal point of the Gaussian beam is set at the origin, where the waist becomes minimum $w(0)=w_0$.
In a uniform material or a vacuum, there is no mechanism to confine the mode, and the beam waist can be arbitrarily controlled by the use of an optical lens up to the diffraction limit.
Thus, $w_0$, and consequently $z_0$, can be controlled and determined by a boundary condition.
It is also useful to note that $kww^{'}=2z/z_0$ and $kw^2/q=2\left(i+z/z_0\right)$ hold, and we obtain
\begin{eqnarray}
&&-kw^2\theta^{'}
-2m \nonumber \\
&&
-4
\left(
  \frac{r}{w}
\right)^2
\frac{f^{'}}{f}
+
2(m+1)
\frac{f^{'}}{f}
+2
\left(
  \frac{r}{w}
\right)^2
\frac{f^{''}}{f}
=0. 
\end{eqnarray}
We now focus on the last three terms of this equation, which can be rewritten by exchanging valuables subsequently using $\rho=r/w$, $a=\rho^2$, and $b=2a$ as
\begin{eqnarray}
&&-4
\rho^2
\frac{f^{'}}{f}
+
2(m+1)
\frac{f^{'}}{f}
+2
\rho^2
\frac{f^{''}}{f}\nonumber  \\
&=&
\frac{2}{f}
\left[
-2a
\frac{d}{da}
+(m+1)
\frac{d}{da}
+a
\frac{d^2}{da^2}
\right]
f \nonumber  \\
&=&
\frac{4}{f}
\left[
b
\frac{d^2}{db^2}
+(m+1-b)
\frac{d}{db}
\right]
f
=
-4p, 
\end{eqnarray}
where we used the differential equation for the associate Laguerre function, $L_p^m$, (Appendices)
\begin{eqnarray}
\left[
b
\frac{d^2}{db^2}
+(m+1-b)
\frac{d}{db}
\right]
L_p^m (b)
=
-pL_p^m (b),
\end{eqnarray}
such that we can obtain $f(b)=L_p^m (b)
=
L_p^m 
\left(
2
\left(
  \frac{r}{w}
\right)^2
\right)$. 
The rest of the Helmholtz equation is 
\begin{eqnarray}
kw^2 \theta^{'}
=
-2(2p+m),
\end{eqnarray}
which gives the phase-shift
\begin{eqnarray}
\theta
=
-(2p+m)
\tan^{-1}
(z/z_0).
\end{eqnarray}
Finally, we obtain 
\begin{eqnarray}
\psi(r,\phi,z)
=&&
\frac{w_0}{w}
\left(
\frac{\sqrt{2}r}{w}
\right)^{m}
L_p^m 
\left(
2
\left(
  \frac{r}{w}
\right)^2 
\right) \nonumber \\
&&
{\rm  e}^{-\frac{r^2}{w^2}}
{\rm  e}^{ik\frac{r^2}{2 R}}
{\rm  e}^{im\phi}
{\rm  e}^{-i(2p+m+1)\tan^{-1}(z/z_0)},
\end{eqnarray}
which is not normalised, yet.
The norm $N_{\rm norm}$ of the wavefunction is obtained by
\begin{eqnarray}
N_{\rm norm}^2&=&
\int_{0}^{\infty}dr
2\pi r
|\psi(r,\phi,z)|^{2} \nonumber \\
&=&
\frac{\pi}{2}w_0^2
\int_{0}^{\infty}db
b^{m}
\left|
L_p^m 
\left(
b 
\right) 
\right|^2
{\rm  e}^{-b} \nonumber \\
&=&
\frac{\pi}{2}w_0^2
\frac{(p+m)!}{p!}, 
\end{eqnarray}
where we used the orthogonality condition (Appendices)
\begin{eqnarray}
\int_{0}^{\infty}
db
{\rm e}^{-b}
b^{m}
L_l^m (b)
L_n^m (b)
=
\frac{(l+m)!}{l!}
\delta_{l,n}.
\end{eqnarray}
Thus, we obtain
\begin{eqnarray}
N_{\rm norm}=
w_0
\sqrt{
\frac{\pi}{2}w_0^2
\frac{(p+m)!}{p!}
}.
\end{eqnarray}

Now, we consider the case for a negative value of $m$.
The only source of the azimuthal dependence in the Helmholtz equation is coming from 
\begin{eqnarray}
\partial_{\phi} ^2 \psi
=
-\frac{m^2}{r^2}
\psi,
\end{eqnarray}
such that the solution does not depend on the sign of $m$.

Therefore, the final normalised wavefunction becomes
\begin{eqnarray}
\psi(r,\phi,z)
=&
\sqrt{
\frac{2}{\pi}
\frac{p!}{(p+|m|)!}
}
\frac{1}{w}
\left(
\frac{\sqrt{2}r}{w}
\right)^{|m|}
L_p^{|m|} 
\left(
2
\left(
 \frac{r}{w}
\right)^2 
\right) \nonumber \\
&
{\rm  e}^{-\frac{r^2}{w^2}}
{\rm  e}^{ik\frac{r^2}{2 R}}
{\rm  e}^{im\phi}
{\rm  e}^{-i(2p+|m|+1)\tan^{-1}(z/z_0)}. \nonumber \\
\end{eqnarray}
Here, the wavefunction was normalised in the $xy$-plane as
\begin{eqnarray}
\int_{0}^{\infty}dr
2\pi r
|\psi(r,\phi,z)|^{2}
&=&
1,
\end{eqnarray}
since our main interests in the following sections are orbital angular momentum, and this normalisation is easier to treat.
On the other hand, in the consideration of the electric field in the previous subsection, the normalisation was slightly different, since we have prioritised to have the proper definition of $E_0$ (V/cm) as the electric field.
When the number of photons that we are considering is one, this corresponds to the zero-point fluctuation of the electric field, $e_0=\sqrt{2 \hbar \omega /\epsilon V}$, but actually a laser beam contains a macroscopic number of photons.
By comparing the factors between $E_0$ and $\psi(r,\phi,z)$, we realise that we should assume $V=w(z)^2L$, where $L$ is the length of the system along $z$.
This means that the magnitude of the electric field changes upon propagation due to the change of the beam waist. 
If the beam expands, the electric field decreases, and {\it vice versa}.
This is attributed to the change in size of the mode profile for photons.
If we use the normalisation of this subsection, $E_0$ should be simply re-defined as $E_0
=\sqrt{
2 \hbar \omega N/\epsilon L
  }$.

It is worth making a remark on the Gouy phase \cite{Pancharatnam56,Berry84,Tomita86,Allen92,Simon93,Hamazaki06,Bliokh09} of  
\begin{eqnarray}
\phi_{\rm G}=(2p+|m|+1)\tan^{-1}(z/z_0),
\end{eqnarray}
which is the same as a geometrical phase of Pancharatnam-Berry.
This term appears due to the focusing of the beam, which will change $(\mathcal{E}_x,\mathcal{E}_y)$ at $z\rightarrow -\infty$ to $(-\mathcal{E}_x,-\mathcal{E}_y)$ at $z\rightarrow +\infty$ upon crossing the focal point at $z=0$.
This change is taken into account within our orbital wavefunction $\psi(r,\phi,z)$, where the polarisation state is not changed.
In the absence of OAM, corresponding to $m=0$ and $p=0$, this change just accompanies a phase-shift of $\pi$.
With OAM, the extra phase factor of ${\rm e}^{im\phi}$ will contribute to it as an additional phase-shift of $\pi m$, because the focusing corresponds to rotating the phase from $\phi=0$ to $\phi=\pi$.
This global change of the orbital due to focussing together with the local rotation of the phase by OAM is responsible for the Gouy phase.
In addition, the radial distribution due to the mode shape described by a Laguerre function will also contribute, in a similar way. 
For the radial profile, we have $p$-nodes along the radial direction, where the focussing corresponds to change the phase front located at $r=w(z)$ to $r=-w(z)$.
During this change, the nodes along $r$ will go across the origin, while adding a phase-shift of $\pi p$. 
In addition, there are nodes at $-r$, or equivalently $(r,\phi=-\pi)$, along the opposite radial direction.
Therefore, the total contribution to the phase-shift from radial oscillation is $2\pi p$. 
The actual change of the phase is not abrupt, and it adiabatically changes in the length scale of $z_0$.
For the propagation in a GRIN fibre, this Gouy phase is, fortunately, not so important because of the absence of focusing, as we shall see in the next subsection.

We should be careful for the interpretation of the radial quantum number, $p$, which describes the number of nodes along the radial direction.
This value is different from the quantum number, $l$, to describe the magnitude of OAM in a spherical symmetric system by $Y_l^m$, for which the value of $m$ is limited to be $m=l, l-1, \cdots, 0, \cdots -(l-1),-l$.
On the other hand, there is no such restriction to $L_p^{|m|}$, where the so-called LG$_{01}$ mode at $p=0$ and $m=1$ can be well-defined, for example.
This also means that $p$ cannot be the proper quantum number to be assigned as the magnitude of OAM.
In fact, the LG mode is not the simultaneous eigenstate of the magnitude of OAM and the component of OAM along the quantised axis ($z$), albeit the expectation value of the magnitude depending on $p$.
It is reasonable to expect that there exists the simultaneous eigenstates, according to the general theory of OAM, but LG modes do not diagonalise the operator for the magnitude of OAM.
Therefore, we also use $n$ for the radial quantum number later, instead of the popular use of $p$ to avoid unnecessary confusion to the momentum, $p=\hbar k$.

\subsection{Laguerre-Gauss mode in a graded index fibre}
As another example, for which the Helmholtz equation can be solved exactly, we also discuss the propagation of a coherent monochromatic laser beam in the GRIN fibre \cite{Kawakami68,Yariv97}, which has the refractive index $n(r)$ dependence given by 
$n(r)^2
=\epsilon(r)/\epsilon_0
=
n_0^2
\left(
1-g^2 r^2
\right)$, where the graded index parameter, $g=2\pi/ \Lambda$, has the dimension of inverse length, and $g$ must be small to justify the derivation of the Helmholtz equation such that the index profile is sufficiently gentle ($\Lambda \gg \lambda$).
We also define the wavenumber in a vacuum as $k_0=2\pi/ \lambda$ for a laser beam emitted from the waveguide.
The energy of the photon will not be changed upon the emission, such that $\omega=ck_0$ is valid, while the dispersion relationship, $\omega=\omega(k)$, in the waveguide is highly non-trivial, and we will obtain this from the Helmholtz equation.
We also define a constant wavenumber parameter, 
$k_{n_0}=2 \pi/ \lambda_{n_0}
=\frac{2 \pi n_0}{\lambda}
=k_0 n_0$,
since the waveguide is mostly determined by the core refractive index, $n_0=n(0)$, and this is simply a constant parameter, where $k_{n_0}$ is different from the true wavenumber, $k$, responsible for describing photon momentum of $p=\hbar k$.

With those parameters, we can rewrite 
$
\mu_0
\epsilon
\omega^2
=
n(r)^2/c^2
(c k_0)^2
=
k_0^2
n(r)^2
=
k_{n_0}^2(1-g^2r^2)
$,
and the Helmholtz equation becomes
\begin{eqnarray}
\left(
\nabla^2
+2ik
\frac{\partial }{\partial z}
+
(k_{n_0}^2-k^2)
+k_{n_0}^2 g^2 r^2
\right)
\psi
=0.
\end{eqnarray}
In the cylindrical coordinate, we can convert this equation to 
\begin{eqnarray}
&&\left(
  \partial_r^2  
+
\frac{1}{r} 
\partial_r 
+2ik
\partial z
+
\partial z^2
\right. \nonumber \\
&&\left.
+
\frac{1}{r^2} 
\partial_{\phi}^2 
+
(k_{n_0}^2-k^2)
+k_{n_0}^2 g^2 r^2
\right)
\psi
=0
\label{eq51}
\end{eqnarray}

One of the conceptual advantages to consider the GRIN waveguide is that we do not have to worry about the paraxial slowly varying approximation at all, because the second derivative along $z$ vanishes.
This can be verified by confirming that the trial wavefunction
\begin{eqnarray}
\psi(r,\phi,z)
&=
\left (
  \frac{r}{w_0}
\right)^{m}
f
\left(
\left(
  \frac{r}{w_0}
\right)^{2}
\right)
{\rm  e}^{ik_{n_{0}}\frac{r^2}{2 q}+im\phi},
\label{eq52}
\end{eqnarray}
with the constant waist $w_0$ and the constant complex radius, $q$, becomes the solution.
Again, we tentatively assume $m \ge 0$.
By inserting Eq.~(\ref{eq52}) into Eq.~(\ref{eq51}), we obtain
\begin{eqnarray}
&
2k_{n_{0}}i
\frac{1}{q}
+
k_{n_{0}}^2 r^2
\left(
-\frac{1}{q^2}
+g^2
\right)
+
\left(
k_{n_{0}}^2 
-k^2
\right)
\nonumber \\
&
+\frac{4}{w_0^2}
\left(
  \frac{r}{w_0}
\right)^2
\frac{f^{''}}{f}
+
\frac{4k_{n_{0}}i}{q}
\left(
  \frac{r}{w_0}
\right)^2
\frac{f^{'}}{f}
+
\frac{2(2m+1)}{w^2}
\frac{f^{'}}{f}
+
\frac{2}{w_0^2}
\frac{f^{'}}{f} \nonumber \\
&
+
2k_{n_{0}}i
\frac{m}{q}
=0.
\end{eqnarray}
We then obtain a stable Gaussian form by noting
\begin{eqnarray}
q&=&- i\frac{i}{g} \\
w_0&=&
\sqrt{
  \frac{2}{gk_{n_0}}
}.
\end{eqnarray}
We also use useful identities
\begin{eqnarray}
2k_{n_0}i
\frac{1}{q}
&=&
-4
\frac{4}{w_0^2} \\
g w_0^2
&=&
  \frac{2}{k_{n_0}},
\end{eqnarray}
and the Helmholtz equation then becomes
\begin{eqnarray}
&&
-2(m+1)
+
(
k_{n_0}^2
-k^2
)
\frac{w_0^2}{2} \nonumber \\
&+&
\left[
2
\left(
\frac{r}{w_0} 
\right)^2
\frac{f^{''}}{f}
+2
(m+1)
\frac{f^{'}}{f}
-4
\left(
\frac{r}{w_0} 
\right)^2
\frac{f^{'}}{f}
  \right ] 
=0. \nonumber \\
\label{eq58}
\end{eqnarray}
By noticing that the last three terms in the left-hand side of Eq.~({\ref{eq58}) can be described by the associated Laguerre function, we obtain
\begin{eqnarray}
f
=
L_n^m 
\left(
2
\left(
  \frac{r}{w_0}
\right)^2
\right),
\end{eqnarray}
and the remaining equation becomes
\begin{eqnarray}
k^2
=
k_{n_0}^2
-
2g k_{n_0} 
(2n+m+1).
\end{eqnarray}
This provides the solution \cite{Yariv97}
\begin{eqnarray}
k
=
\frac{\omega}{v_0}
\sqrt{
1
-
\frac{\delta \omega_0}{\omega} 
(2n+m+1)
},
\label{eq61}
\end{eqnarray}
where we have defined a phase velocity at the core as $v_0=c/n_0$ and a frequency shift as $\delta \omega_{0}=v_0 g$.
By solving Eq.~(\ref{eq61}) with respect to $\omega$, we obtain the dispersion relationship
\begin{eqnarray}
\omega
&=
\sqrt{v_0^2 k^2+ \delta \omega_0^2 (2n+m+1)^2}
+
\delta \omega_0 (2n+m+1) . \nonumber \\
\end{eqnarray}
This dispersion relationship can be intuitively understood as follows: Since $\omega \neq 0$ at $k=0$, this implies an opening of an energy gap in a band diagram, meaning that the dispersion is '{\it massive}'. 
The emergence of an energy gap is reminiscent of the theory of superconductivity \cite{Schrieffer71} and the Nambu-Anderson-Goldstone-Higgs theory of a broken symmetry \cite{Nambu59,Anderson58,Goldstone62,Higgs64}.
We infer that a similar symmetry principle is hidden in our system.  
We will discuss this in a subsequent paper.
Here, we can recognise the increase of the gap by increasing the radial quantum number, $p$, and the quantised OAM number, $m$, because the discrete photon energy is related to the confinement degrees of freedom of photons rather than the free propagation of photons along $z$.

Finally, we relax the condition for $m$ to allow negative integers without breaking the formalism.  Normalising the wavefunction as before, we obtain an exact solution
\begin{eqnarray}
\psi(r,\phi,z)
=&&
\sqrt{
\frac{2}{\pi}
\frac{n!}{(n+|m|)!}
}
\frac{1}{w_0}
\left(
\frac{\sqrt{2}r}{w_0}
\right)^{|m|} \nonumber\\
&& \ \ 
L_n^{|m|} 
\left(
2
\left(
  \frac{r}{w_0}
\right)^2 
\right) 
{\rm  e}^{-\frac{r^2}{w_0^2}}
{\rm  e}^{im\phi},
\end{eqnarray}
in a GRIN fibre without the slowly varying paraxial approximation.

\section{Methods}
\subsection{Orbital Angular Momentum for Photons}
We have confirmed the fundamental principle on how to treat a coherent laser beam on the basis of Maxwell's equations and a quantum many-body theory.
In particular, we have understood why we can describe a macroscopically coherent laser by a single-particle wavefunction, $\Psi({\bf r})$, due to Bose-Einstein statistics, while the entire many-body state is described by a coherent state with both orbital and spin degrees of freedom.
Photons are quantum-mechanical particles with a wave nature; we can also describe them with Maxwell's equations together with a quantum many-body theory.
For coherent photons from a laser, it was less obvious how we can treat the ray quantum-mechanically; however, a laser produces indistinguishable photons with the same phase by the stimulated emission process, in which existing photons in a cavity induce recombinations of electron-hole pairs to make clones of photons as a results of a chain-reaction.
Thus, we can describe a coherent monochromatic ray by the single mode of $\Psi({\bf r})$.
If the waveguide contains several modes, it is straightforward to allow the superposition of these macroscopically coherent rays.

The fundamental equation for describing the orbital character of $\Psi({\bf r})$ is the Helmholtz equation, instead of the Schr\"odinger equation, although we have a significant similarity to a paraxial wave (Table \ref{tab:table1}).
Unlike in a vacuum without a material, where $\Psi({\bf r})$ is a simple plane-wave, the mode profile of $\Psi({\bf r})$ can be highly non-trivial in materials, depending on the symmetries of the waveguides and the actual profile of the refractive index, $n({\bf r})$.
In the previous section, we have obtained LG modes in a GRIN fibre with a uniform material. Here, the LG modes (LG$_{nm}$) were clearly labelled by the radial quantum number $n$ and the quantum optical orbital angular momentum number $m$ along the propagation direction.
The central theme of this paper is to examine the validity of this interpretation that $m$ is indeed a proper quantum index to describe the optical OAM, and thus the angular momentum of the orbital is $\hbar m$.
Under the assumption that a standard quantum mechanical treatment is also applicable to photons, described by $\Psi({\bf r})$, we examine the impacts of OAM operations in the following sections.

\subsection{Canonical orbital angular momentum operators}
The most well-established quantum mechanical treatment is canonical commutation  relationship for the position $\hat{\bf r}=(\hat{x},\hat{y},\hat{z})$ and the momentum $\hat{\bf p}=(\hat{p}_x,\hat{p}_y,\hat{p}_z)$ operators \cite{Dirac30,Baym69,Sakurai14}: $[\hat{x},\hat{p}_x]=i\hbar$, $[\hat{y},\hat{p}_y]=i\hbar$, $[\hat{z},\hat{p}_z]=i\hbar$, and commutable relationship among other combinations.
The OAM operator, $\hat{\bf l}$, is defined by use of those $\hat{\bf r}$ and $\hat{\bf p}$ as $\hat{\bf l}=\hat{\bf r}\times \hat{\bf p}$, such that each component becomes
\begin{eqnarray}
\hat{l}_x
&=&
\frac{\hbar}{i}
\left(
y \partial_z
-
z \partial_y
\right) \\
\hat{l}_y
&=&
\frac{\hbar}{i}
\left(
z \partial_x
-
x \partial_z
\right) \\
\hat{l}_z
&=&
\frac{\hbar}{i}
\left(
x \partial_y
-
y \partial_x
\right),
\end{eqnarray} 
respectively.
In a system with a spherical symmetry, the eigenstate of these operators are described by $Y_{lm}(\theta,\phi)=\langle \theta,\phi | l,m \rangle$,  where $\theta$ and $\phi$ are polar and azimuthal angels, respectively, $l$ and $m$ are quantum numbers for the magnitude of OAM and the OAM component along the quantisation axis, respectively \cite{Dirac30, Baym69,Sakurai14}. 
Our goal is to obtain a similar relationship for a system with a cylindrical symmetry, described by the LG modes.
In this section, we will obtain the operator representation of $\hat{\bf l}$ in a cylindrical coordinate.

The unit vectors of a cylindrical coordinate are defined for a rotation in a $3D$ Cartesian coordinate along the $z$ axis, 
\begin{eqnarray}
\left (
  \begin{array}{c}
   \hat{\bf r}
\\
   \hat{\bf \Phi}
  \end{array}
\right)
= 
\left (
  \begin{array}{cc}
   \cos \phi & \sin \phi
\\
   -\sin \phi & \cos \phi
  \end{array}
\right)
\left (
  \begin{array}{c}
   \hat{\bf x}
\\
   \hat{\bf y}
  \end{array}
\right),
\end{eqnarray}
where $\hat{\bf r}$ is the unit vector along $r$  and $\hat{\bf \Phi}$ is a unit vector along the azimuthal direction, while $z$ is unchanged.
The important point in this coordinate is $\phi$ dependence of these unit vectors, i.e., $\hat{\bf r}=\hat{\bf r}(\phi)$ and $\hat{\bf \Phi}=\hat{\bf \Phi}(\phi)$.
Therefore, the nabla operator is of the form
\begin{eqnarray}
\nabla
=
\left(
  \frac{\partial}{\partial r},
  \frac{1}{r}\frac{\partial}{\phi},
  \frac{\partial}{\partial z}
\right)
=
\frac{\partial}{\partial r}
\hat{\bf r} 
+\frac{1}{r}\frac{\partial}{\partial \phi}
\hat{\bf \Phi} 
+\frac{\partial}{\partial z}
\hat{\bf z}, 
\end{eqnarray}
and the Laplacian  becomes
\begin{eqnarray}
\nabla^2
=
\frac{\partial^2}{\partial r^2}
+\frac{\partial^2}{\partial z^2}
+\frac{1}{r^2}\frac{\partial^2}{\partial \phi^2}
+
\frac{1}{r}
\frac{\partial}{\partial r},
\end{eqnarray}
where we have used 
\begin{eqnarray}
\frac{\partial}{\partial \phi}
 \hat{\bf r} (\phi) 
=
\hat{\bf \Phi}.
\end{eqnarray}
This $\phi$ dependence of $\hat{\bf r}(\phi)$ and $\hat{\bf \Phi}(\phi)$ makes it difficult to define and treat the OAM operators for $\hat{l}_r$ and $\hat{l}_\phi$.
Instead, we will keep using $\hat{l}_x$ and $\hat{l}_y$, defined above, and we will express them by $(r,\phi,z)$.
In order to use $(r,\phi)$ instead of $(x,y)$, we obtain
\begin{eqnarray}
\partial_x
&=&
\cos \phi
\partial_r
-\frac{1}{r}
\sin \phi
\partial_\phi, \\
\partial_y
&=&
\sin \phi
\partial_r
+\frac{1}{r}
\cos \phi
\partial_\phi .
\end{eqnarray}
By inserting these into $\hat{\bf l}$, we obtain
\begin{eqnarray}
\hat{l}_x
&=&
\frac{\hbar}{i}
\left(
-z \sin \phi \partial_r
-
\frac{z}{r}
\cos \phi \partial_{\phi}
+
r \sin \phi
\partial_z
\right), \\
\hat{l}_y
&=&
\frac{\hbar}{i}
\left(
z \cos \phi \partial_r
-
\frac{z}{r}
\sin \phi \partial_{\phi}
-
r \cos \phi
\partial_z
\right), \\
\hat{l}_z
&=&
\frac{\hbar}{i}
\partial_{\phi} .
\end{eqnarray}

We can readily confirm the original commutation relationship
\begin{eqnarray}
\left [
\hat{l}_x,\hat{l}_y
\right]
=
i \hbar
\hat{l}_z
\end{eqnarray}
and its cyclic exchanges
\begin{eqnarray}
\left [
\hat{l}_y,\hat{l}_z
\right]
&=&
i \hbar
\hat{l}_x, \\
\left [
\hat{l}_z,\hat{l}_x
\right]
&=&
i \hbar
\hat{l}_y
\end{eqnarray}
are all valid in the cylindrical coordinate $(r,\phi,z)$.

We also obtain raising and lowering operators, respectively, as
\begin{eqnarray}
\hat{l}_{+}
&=&
\hat{l}_x
+
i
\hat{l}_y \\
&=&
\hbar
{\rm e}^{i \phi}
\left(
z \partial_r
+
\frac{z}{r}
i
\partial_{\phi}
-
r 
\partial_z
\right), \\
\hat{l}_{-}
&=&
\hat{l}_x
-
i
\hat{l}_y \\
&=&
\hbar
{\rm e}^{-i \phi}
\left(
- z \partial_r
+
\frac{z}{r}
i
\partial_{\phi}
+
r 
\partial_z
\right) .
\end{eqnarray}

\subsection{Application to plane waves \& problems}
So far, it was straightforward to develop a theory for photonic OAM. 
In this subsection, we will apply our canonical OAM operator to plane waves to clarify that problems arise.
Specifically, we consider a plane wave with OAM in the simplest form:
\begin{eqnarray}
\Psi(r,\phi,z)
=&
{\rm e}^{i k z}
{\rm e}^{i m \phi},
\end{eqnarray}
which is {\it not} the solution of the Helmholtz equation at $m \neq 0$.
Nevertheless, it is useful to clarify the potential issue and to explain what we should address in the following sections.

First, multiplying it by $\hat{\bf l}$, we obtain
\begin{eqnarray}
\hat{l}_{+}
(r,\phi,z)
\Psi
&=&
\hbar
{\rm e}^{i  \phi}
\left(
-\frac{z}{r}
m
-
i
k
r 
\right)
\Psi, \\
\hat{l}_{-}
(r,\phi,z)
\Psi
&=&
\hbar
{\rm e}^{-i \phi}
\left(
-\frac{z}{r}
m
+
i
k
r 
\right)
\Psi, \\
\hat{l}_{z}
(r,\phi,z)
\Psi
&=&
\hbar m
\Psi, 
\end{eqnarray}
which means that $\Psi$ is indeed an eigenstate for $l_z$ and that $\hat{l}_{\pm}$ is effectively working to raise and lower the eigenvalue of the angular momentum component along the direction of the propagation.
If we multiply them by $\Psi^{*}$ from the left, we obtain
\begin{eqnarray}
\Psi^{*} \ 
\hat{l}_{+}
(r,\phi,z)
\ \Psi
&=&
\hbar
{\rm e}^{ i \phi}
\left(
-\frac{z}{r}
m
-
i
k
r 
\right), \\
\Psi^{*} \ 
\hat{l}_{-}
(r,\phi,z) \ 
\Psi
&=&
\hbar
{\rm e}^{- i \phi}
\left(
-\frac{z}{r}
m
+
i
k
r 
\right), \\
\Psi^{*} \hat{l}_{z}
(r,\phi,z)
\Psi
&=&
\hbar m.
\end{eqnarray}
By averaging these over space, we obtain
\begin{eqnarray}
\left \langle
\hat{l}_{x}
(r,\phi,z)
\right \rangle
&=&
0, \\
\left \langle
\hat{l}_{y}
(r,\phi,z)
\right \rangle
&=&
0, \\
\left \langle
\hat{l}_{z}
(r,\phi,z)
\right \rangle
&=&
\hbar m . 
\end{eqnarray}
Therefore, the expectation values are reasonably well-defined.

However, if we calculate the complex conjugate of $\hat{l}_{-} \Psi$ simply by taking its complex conjugate as 
\begin{eqnarray}
\Psi^{*}
\left(
\hat{l}_{-}
(r,\phi,z)
\right)^{\dagger}
&=
\hbar
{\rm e}^{-i (m-1) \phi}
\left(
-\frac{z}{r}
m
-
i
k
r 
\right)
{\rm e}^{-i k z}, \nonumber \\
\end{eqnarray}
and multiply this by $\hat{l}_{-} \Psi$ from the right to calculate the norm, we obtain
\begin{eqnarray}
\Psi^{*}
\left(
\hat{l}_{-}
(r,\phi,z)
\right)^{\dagger}
\hat{l}_{-}
(r,\phi,z)
\Psi
&=&
\hbar^2
\left(
\frac{z^2}{r^2}
m^2
+k^2 r^2
\right), \nonumber \\
\end{eqnarray}
which is a positive real value.
On the other hand, the direct calculation of $\Psi^{*} \hat{l}_{+} \hat{l}_{-} \Psi$ becomes
\begin{eqnarray}
\Psi^{*}
\hat{l}_{+}
(r,\phi,z)
\hat{l}_{-}
(r,\phi,z)
\Psi
&=
\hbar^2
\left(
\frac{z^2}{r^2}m^2
+2ikz
+(kr)^2
+m
\right). \nonumber \\
\end{eqnarray}
This implies that 
\begin{eqnarray}
\left(
\hat{l}_{-}
(r,\phi,z)
\right)^{\dagger}
\neq
\hat{l}_{+}
(r,\phi,z),
\end{eqnarray}
which means that the $l_{\pm}$ is not observable for the Hilbert space spanned by the plane waves with OAM.
We can also confirm the conjugate relationships
\begin{eqnarray}
\Psi^{*}
\left(
\hat{l}_{+}
(r,\phi,z)
\right)^{\dagger}
\hat{l}_{+}
(r,\phi,z)
\Psi
=
\hbar^2
\left(
\frac{z^2}{r^2}
m^2
+k^2 r^2
\right) \nonumber \\ \\
\Psi^{*}
\hat{l}_{-}
(r,\phi,z)
\hat{l}_{+}
(r,\phi,z)
\Psi
=
\hbar^2
\left(
\frac{z^2}{r^2}m^2
+2ikz
+(kr)^2
-m
\right), \nonumber \\
\end{eqnarray}
which also imply
\begin{eqnarray}
\left(
\hat{l}_{+}
(r,\phi,z)
\right)^{\dagger}
\neq
\hat{l}_{-}
(r,\phi,z).
\end{eqnarray}
We also see 
\begin{eqnarray}
&&\Psi^{*}
\left(
\hat{l}_{-}
(r,\phi,z)
\right)^{\dagger}
\hat{l}_{-}
(r,\phi,z)
\Psi \nonumber \\
&&=
\Psi^{*}
\left(
\hat{l}_{+}
(r,\phi,z)
\right)^{\dagger}
\hat{l}_{+}
(r,\phi,z)
\Psi, 
\end{eqnarray}
showing a classical result without providing commutation relationship.
This is a remarkable difference from the standard quantum mechanics \cite{Sakurai14}, which shows the canonical commutation relationship upon the calculation of the norm for $\hat{l}_{\pm} Y_{l}^{m}(\theta,\phi)$.

On the other hand, the direct calculation shows  
\begin{eqnarray}
&\Psi^{*}
\hat{l}_{+}
(r,\phi,z)
\hat{l}_{-}
(r,\phi,z)
\Psi
-
\Psi^{*}
\hat{l}_{-}
(r,\phi,z)
\hat{l}_{+}
(r,\phi,z)
\Psi
\nonumber \\
&=
2
\hbar
\Psi^{*}
\hat{l}_{z}
(r,\phi,z)
\Psi,
\end{eqnarray}
such that the commutation relationship
\begin{eqnarray}
\left[
\hat{l}_{+}
(r,\phi,z)
,
\hat{l}_{-}
(r,\phi,z)
\right]
=
2
\hbar
\hat{l}_{z}
(r,\phi,z)
\end{eqnarray}
is indeed satisfied on the average.

These apparent contradiction and inconsistency are coming from the assumption of the ill-defined plane-wave wavefunction, $\Psi(r,\phi,z)={\rm e}^{i k z}{\rm e}^{i m \phi}$.
This is confirmed by calculating the magnitude of the OAM along the radial direction as
\begin{eqnarray}
\Psi^{*}  
\left(
\hat{l}_x^2
+
\hat{l}_y^2
\right)
\Psi
=&
\hbar^2
\left(
m^2\frac{z^2}{r^2}
+
k^2 r^2
+ 2 i k z 
\right),
\end{eqnarray}
which gives an imaginary part, thus showing that the magnitude is not observable.
If we take the average over $z \in (0,L)$, we obtain
\begin{eqnarray}
\int_{0}^{L}\frac{dz}{L}
\Psi^{*}  
\left(
\hat{l}_x^2
+
\hat{l}_y^2
\right)
\Psi
=
\hbar^2
\left(
m^2\frac{L^2}{3r^2}
+
k^2 r^2
+  i k L 
\right), \nonumber \\
\end{eqnarray}
which is still a complex value.
Further average over $r\in(0,R)$, where $R$ is the radius of a cylindrical waveguide, gives
\begin{eqnarray}
&&
\int_0^{R}\frac{2\pi r dr}{\pi R^2}
\int_{0}^{L} \frac{dz}{L}
\Psi^{*}  
\left(
\hat{l}_x^2
+
\hat{l}_y^2
\right)
\Psi \nonumber \\
&&=
\hbar^2
\left(
\frac{2m^2L^2}{3R^2}(\ln R-\ln 0)
+\frac{1}{2}k^2R^2
+  i k L 
\right) ,
\end{eqnarray}
which diverges at the origin.
We can also integrate over $z \in (L/2,L/2)$, and obtain 

\begin{eqnarray}
&&
\int_0^{R}\frac{2\pi r dr}{\pi R^2}
\int_{-L/2}^{L/2} \frac{dz}{L}
\Psi^{*}  
\left(
\hat{l}_x^2
+
\hat{l}_y^2
\right)
\Psi \nonumber \\
&&=
\hbar^2
\left(
\frac{m^2L^2}{6R^2}(\ln R-\ln 0)
+\frac{1}{2}k^2R^2
\right),
\end{eqnarray}
which becomes a real value, but still diverges at the origin.

The position-dependent average of the radial magnitude suggests that it contains extrinsic contributions of OAM.
For both coordinates, we could not avoid the ultraviolet divergences at the origin, which are coming from the finite amplitude of the wavefunction at the origin.
Without having a node at the origin, the magnitude of the OAM required to sustain the phase described by ${\rm e}^{i m \phi}$ is impossible to exist.

However, for the LG modes, which always have nodes at the centre ($r=0$) of the waveguide for $^{\forall} m \neq0$, there is a chance that the OAM can be well-defined quantum-mechanically.
Our main purpose of this work is to confirm the validity of this concept of OAM, using canonical orbital angular momentum operators defined in this section, for the LG modes in a cylindrical GRIN fibre.
In the next section, we will confirm positive results, including the observable nature of the magnitude and the commutation relationship for OAM.

\section{Results and Discussions}
We consider applications of the canonical OAM operators to the LG modes in a GRIN fibre.
We use the normalised LG mode \cite{Allen92,Enk94,Barnett16,Yariv97,Jackson99}
\begin{eqnarray}
&&\Psi_n^m(r,\phi,z)
=
\langle r,\phi,z | \Psi_n^m \rangle \\
&&=
\frac{1}{w_0}
\sqrt{
\frac{2}{\pi}
\frac{n!}{(n+|m|)!}
}
\left(
\frac{\sqrt{2}r}{w_0}
\right)^{|m|} 
L_n^{|m|} 
\left(
2
\left(
  \frac{r}{w_0}
\right)^2 
\right) 
\nonumber \\
&& \ \  \ \  
{\rm  e}^{-\frac{r^2}{w_0^2}}
{\rm  e}^{i m \phi}
{\rm  e}^{i kz},
\end{eqnarray}
where $n$ is the radial quantum number and $m$ is the quantum number of OAM along the direction of propagation.
In principle, we should also consider a superposition state made of the LG modes with different quantum numbers \cite{Padgett99,Milione11,Liu17}, but we will not consider this in this paper for simplicity; yet our formalism works well.
We define a normalised cross-sectional area as $a=2r^2/w_0^2$ to simplify calculations.
We use various formulas for associate Laguerre functions, which are summarised in Appendixes.

\subsection{Expectation value}
First, we have checked the expectation values of $\hat{\bf l}$ by use of the LG modes.
This was straightforward by noting
$\hat{l}_{\pm}
\psi_{n}^{m}
\propto
\psi_{n}^{m\pm 1}
\propto
{\rm e}^{i (m \pm 1)\phi}$
and 
\begin{eqnarray}
\int_{0}^{2\pi}
\frac{d \phi}{2\pi}
{\rm e}^{\pm i \phi}
=0,
\end{eqnarray}
and thus we obtain
\begin{eqnarray}
\left \langle
\hat{l}_x
\right \rangle
=&
0,
\\
\left \langle
\hat{l}_y
\right \rangle
=&
0,
\\
\left \langle
\hat{l}_z
\right \rangle
=&
\hbar m .
\end{eqnarray}
Therefore, the quantum-mechanical expectation value of OAM is well-defined for all directions.
This is a single particle expectation value, and the total angular momentum, $\hat{L}_z$, along $z$ for a coherent state is obtained by multiplying $N$ as $\langle \hat{L}_z \rangle=\hbar mN$.

\subsection{Ladder operations}
We will evaluate ladder operations to the LG modes.
The calculations are straightforward but tedious, so that we will split them into several sections.

\subsubsection{Rising operation for $m>0$}
We assume $m>0$ and calculate
\begin{eqnarray}
\hat{l}_{+} \Psi_n^m
=&&
\hbar
{\rm e}^{i (m+1) \phi}
{\rm  e}^{i kz}
\frac{1}{w_0}
\sqrt{
\frac{2}{\pi}
\frac{n!}{(n+m)!}
}
\left(
\frac{\sqrt{2}r}{w_0}
\right)^{m}
{\rm  e}^{-\frac{r^2}{w_0^2}}
\nonumber \\
&& \cdot
\left(
z
\left(
\frac{m}{r}
+4\frac{r}{w_0^2}
\frac{d}{da}
-
2\frac{r^2}{w_0^2}
\right) 
-z \frac{m}{r}
-ikr
\right) 
L_n^{m} (a), \nonumber \\
\end{eqnarray}
where the last factor becomes
\begin{eqnarray}
&&\left(
z
\left(
\frac{m}{r}
+4\frac{r}{w_0^2}
\frac{d}{da}
-
2\frac{r^2}{w_0^2}
\right) 
-z \frac{m}{r}
-ikr
\right) 
L_n^{m} (a)
\nonumber \\
&&=
2\sqrt{2}
\frac{z}{w_0}
\left(
\frac{\sqrt{2}r}{w_0}
\right)
\left(
\frac{d}{da}
L_n^{m} (a)
-
\frac{1}{2}
L_n^{m} (a)
-
\frac{ik w_0^2}{4z}
L_n^{m} (a)
\right) 
\nonumber \\
&&=
-
2\sqrt{2}
\frac{z}{w_0}
\left(
\frac{\sqrt{2}r}{w_0}
\right)
\left(
L_n^{m+1} (a)
-\frac{1}{2}
  \left(
1-
\frac{ik w_0^2}{2z}
  \right)
L_n^{m} (a)
\right), \nonumber \\
\end{eqnarray}
where we have used
\begin{eqnarray}
\left[
\frac{d}{da}
-1
\right]
L_{n}^{m} (a)
=
-L_n^{m+1}(a).
\end{eqnarray}

Therefore, we obtain
\begin{eqnarray}
\hat{l}_{+} 
\Psi_n^m(r,\phi,z)
=&
-
2\sqrt{2}
\frac{z}{w_0}
\hbar
\sqrt{n+m+1}
\Psi_n^{m+1}(r,\phi,z)
\nonumber \\
&
\cdot
\left(
1
-\frac{1}{2}
  \left(
1-
\frac{ik w_0^2}{2z}
  \right)
\frac{L_n^{m} (a)}{L_n^{m+1} (a)}
\right).
\end{eqnarray}
The most significant part of this expression is that we confirm $\hat{l}_{+}\Psi_n^m \propto \Psi_n^{m+1}$, which means that the raising operator properly works to increase the quantum number $m$ of OAM.
Unfortunately, the coefficient is not a constant, which depends on both $z$ and $r$ through $a$.
Therefore, the shape of the orbital would be significantly distorted upon the application of $\hat{l}_{+}$.
Nevertheless, the main role of $\hat{l}_{+}$ to increase $m$ was successfully confirmed for $m>0$.

\subsubsection{Rising operation for $m=0$}
We then continue to calculate for $m=0$ as
\begin{eqnarray}
\hat{l}_{+} \Psi_n^0
=&
\hbar
{\rm e}^{i  \phi}
{\rm  e}^{i kz}
\frac{1}{w_0}
\sqrt{
\frac{2}{\pi}
\frac{n!}{(n+1)!}
}
\sqrt{n+1}
{\rm  e}^{-\frac{r^2}{w_0^2}}
\nonumber \\
& \cdot
\left(
z
\left(
4\frac{r}{w_0^2}
\frac{d}{da}
-
2\frac{r}{w_0^2}
\right) 
-ikr
\right) 
L_n^{0} (a),
\end{eqnarray}
where the last factor becomes
\begin{eqnarray}
&&\left(
z
\left(
+4\frac{r}{w_0^2}
\frac{d}{da}
-
2\frac{r}{w_0^2}
\right) 
-ikr
\right) 
L_n^{0} (a)
\nonumber \\
&&=
2\sqrt{2}
\frac{z}{w_0}
\left(
\frac{\sqrt{2}r}{w_0}
\right)
\left(
\frac{d}{da}
L_n^{0} (a)
-
\frac{1}{2}
L_n^{0} (a)
-
\frac{ik w_0^2}{4z}
L_n^{0} (a)
\right) 
\nonumber \\
&&=
-
2\sqrt{2}
\frac{z}{w_0}
\left(
\frac{\sqrt{2}r}{w_0}
\right)
\left(
L_n^{1} (a)
-\frac{1}{2}
  \left(
1-
\frac{ik w_0^2}{2z}
  \right)
L_n^{0} (a)
\right), \nonumber \\
\end{eqnarray}
and thus we obtain
\begin{eqnarray}
\hat{l}_{+} 
\Psi_n^0(r,\phi,z)
=&
-
2\sqrt{2}
\frac{z}{w_0}
\hbar
\sqrt{n+1}
\Psi_n^{1}(r,\phi,z)
\nonumber \\
&
\cdot
\left(
1
-\frac{1}{2}
  \left(
1-
\frac{ik w_0^2}{2z}
  \right)
\frac{L_n^{0} (a)}{L_n^{1} (a)}
\right),
\end{eqnarray}
which is exactly the same expression with that of $m>0$.

\subsubsection{Rising operation for $m<0$}
We obtain a similar result by directly calculating
\begin{eqnarray}
&&\hat{l}_{+} \Psi_n^m
=
\hbar
{\rm e}^{i (m+1) \phi}
{\rm  e}^{i kz}
\frac{1}{w_0}
\sqrt{
\frac{2}{\pi}
\frac{n!}{(n-m)!}
}
\left(
\frac{\sqrt{2}r}{w_0}
\right)^{-m}
{\rm  e}^{-\frac{r^2}{w_0^2}}
\nonumber \\
&& \cdot
\left(
z
\left(
-\frac{m}{r}
+4\frac{r}{w_0^2}
\frac{d}{da}
-
2\frac{r^2}{w_0^2}
\right) 
-z \frac{m}{r}
-ikr
\right) 
L_n^{-m} (a), \nonumber \\
\end{eqnarray}
where the last factor becomes
\begin{eqnarray}
&&
\left(
z
\left(
-\frac{m}{r}
+4\frac{r}{w_0^2}
\frac{d}{da}
-
2\frac{r^2}{w_0^2}
\right) 
-z \frac{m}{r}
-ikr
\right) 
L_n^{-m} (a)
\nonumber \\
&&=
\sqrt{2}
\frac{z}{w_0}
\left(
\frac{w_0}{\sqrt{2}r}
\right) \nonumber \\
&&
\left(
   2
  \left(
\frac{d}{da}
  -m
  \right)
L_n^{-m} (a)
-
aL_n^{-m} (a)
-
\frac{ik w_0^2}{z}
L_n^{-m} (a)
\right) 
\nonumber \\
&&=
2\sqrt{2}
\frac{z}{w_0}
(n-m)
\left(
\frac{w_0}{\sqrt{2}r}
\right) 
L_n^{-m-1} (a) \nonumber \\
&&\left(
1-
\frac{a}{2}
\frac{1}{n-m}
  \left(
1+
\frac{ik w_0^2}{2z}
  \right)
\frac{L_n^{|m|} (a)}{L_n^{|m+1|} (a)}
\right),
\end{eqnarray}
where we have used
\begin{eqnarray}
\left[
a
\frac{d}{da}
-m
\right]
L_{n}^{-m} (a)
=
(n-m)
L_n^{-m-1}(a).
\end{eqnarray}
Thus, we obtain
\begin{eqnarray}
\hat{l}_{+} 
\Psi_n^m(r,\phi,z)
=&&
2\sqrt{2}
\frac{z}{w_0}
\hbar
\sqrt{n-m}
\Psi_n^{m+1}(r,\phi,z)
\nonumber \\
&&
\cdot
\left(
1
-\frac{a}{2}
\frac{1}{n-m}
  \left(
1
+
\frac{ik w_0^2}{2z}
  \right)
\frac{L_n^{|m|} (a)}{L_n^{|m+1|} (a)}
\right). \nonumber \\
\end{eqnarray}
Therefore, the raising operator is successfully working to increment $m$, independently of the value and the sign of $m$.

\subsubsection{Lowering operation for $m<0$}
Next, we apply the lowering operator to the case for $m<0$, and obtain
\begin{eqnarray}
\hat{l}_{-} \Psi_n^m
&&=
\hbar
{\rm e}^{i (m-1) \phi}
{\rm  e}^{i kz}
\frac{1}{w_0}
\sqrt{
\frac{2}{\pi}
\frac{n!}{(n-m+1)!}
} \nonumber \\
&& 
\sqrt{n-m+1}
\left(
\frac{\sqrt{2}r}{w_0}
\right)^{-m}
\nonumber \\
&& \cdot
{\rm  e}^{-\frac{r^2}{w_0^2}}
\left(
-z
\left(
-\frac{m}{r}
+4\frac{r}{w_0^2}
\frac{d}{da}
-
2\frac{r^2}{w_0^2}
\right) 
\right. \nonumber \\
&& 
\left.
-z \frac{m}{r}
+ikr
\right) 
L_n^{-m} (a), \nonumber \\
\end{eqnarray}
where the last factor becomes
\begin{eqnarray}
&& \left(
-z
\left(
-\frac{m}{r}
+4\frac{r}{w_0^2}
\frac{d}{da}
-
2\frac{r^2}{w_0^2}
\right) 
-z \frac{m}{r}
+ikr
\right) 
L_n^{-m} (a)
\nonumber \\
&=&
-
4z\frac{r}{w_0^2}
\left(
  \left(
\frac{d}{da}
  -
\frac{1}{2}
  \right)
L_n^{-m} (a)
-
aL_n^{m} (a)
-
\frac{ik w_0^2}{4z}
L_n^{-m} (a)
\right) 
\nonumber \\
&=&
2\sqrt{2}z\frac{\sqrt{2}r}{w_0^2}
L_n^{-m+1} (a)
\left(
1-
\frac{1}{2}
  \left(
1-
\frac{ik w_0^2}{2z}
  \right)
\frac{L_n^{|m|} (a)}{L_n^{|m-1|} (a)}
\right) ,\nonumber \\
\end{eqnarray}
which yields
\begin{eqnarray}
\hat{l}_{-} \Psi_n^m(r,\phi,z)
=&&
2\sqrt{2}
\frac{z}{w_0}
\hbar
\sqrt{n-m+1} \, 
\Psi_n^{m-1}(r,\phi,z)
\nonumber \\
&&
\cdot
\left(
1-
\frac{1}{2}
  \left(
1-
\frac{ik w_0^2}{2z}
  \right)
\frac{L_n^{|m|} (a)}{L_n^{|m-1|} (a)}
\right) . \nonumber \\
\end{eqnarray}
This also shows that the lowering operator can actually lower $m$ to $m-1$.

\subsubsection{Lowering operation for $m=0$}
Similarly, we obtain for $m=0$:
\begin{eqnarray}
\hat{l}_{-} \Psi_n^0
=&
\hbar
{\rm e}^{-i  \phi}
{\rm  e}^{i kz}
\frac{1}{w_0}
\sqrt{
\frac{2}{\pi}
\frac{n!}{(n+1)!}
}
\sqrt{n+1}
{\rm  e}^{-\frac{r^2}{w_0^2}}
\nonumber \\
& \cdot
\left(
-z
\left(
4\frac{r}{w_0^2}
\frac{d}{da}
-
2\frac{r}{w_0^2}
\right) 
+ikr
\right) 
L_n^{0} (a),
\end{eqnarray}
where the last factor becomes
\begin{eqnarray}
&&\left(
-z
\left(
4\frac{r}{w_0^2}
\frac{d}{da}
-
2\frac{r}{w_0^2}
\right) 
-ikr
\right) 
L_n^{0} (a)
\nonumber \\
&&=
-
2\sqrt{2}
\frac{\sqrt{2}r}{w_0}
\left(
L_n^{1} (a)
-\frac{1}{2}
  \left(
1-
\frac{ik w_0^2}{2z}
  \right)
L_n^{0} (a)
\right),
\end{eqnarray}
which yields
\begin{eqnarray}
\hat{l}_{-} \Psi_n^0(r,\phi,z)
=&&
-
2\sqrt{2}
\frac{z}{w_0}
\hbar
\sqrt{n+1}
\Psi_n^{-1}(r,\phi,z) \nonumber \\
&&
\left(
1
-\frac{1}{2}
  \left(
1-
\frac{ik w_0^2}{2z}
  \right)
\frac{L_n^{0} (a)}{L_n^{1} (a)}
\right).
\end{eqnarray}
This is the same formula as that we obtained for $m<0$.

\subsubsection{Lowering operation for $m>0$}
Finally, we apply the lowering operator to the case
for $m>0$, and obtain
\begin{eqnarray}
&&\hat{l}_{-} \Psi_n^m  \nonumber \\
&&=
\hbar
{\rm e}^{i (m-1) \phi}
{\rm  e}^{i kz}
\frac{1}{w_0}
\sqrt{
\frac{2}{\pi}
\frac{n!}{(n+m-1)!}
}
\frac{1}{\sqrt{n+m}} 
\left(
\frac{\sqrt{2}r}{w_0}
\right)^{m}
\nonumber \\
&& \cdot
{\rm  e}^{-\frac{r^2}{w_0^2}}
\left(
-
z
\left(
\frac{m}{r}
+4\frac{r}{w_0^2}
\frac{d}{da}
-
2\frac{r^2}{w_0^2}
\right) 
-z \frac{m}{r}
+ikr
\right) 
L_n^{m} (a), \nonumber \\
\end{eqnarray}
where the last factor becomes 
\begin{eqnarray}
&&\left(
-z
\left(
\frac{m}{r}
+4\frac{r}{w_0^2}
\frac{d}{da}
-
2\frac{r^2}{w_0^2}
\right) 
-z \frac{m}{r}
+ikr
\right) 
L_n^{m} (a)
\nonumber \\
&&=
-\frac{z}{r}
\left(
2
\left(
a 
\frac{d}{da}
+m
\right)
-a
-
\frac{ik r^2}{z}
\right) 
L_n^{m} (a)
\nonumber \\
&&=
-
2
\frac{z}{r}
\left(
(n+m)
L_n^{m-1} (a)
-\frac{r^2}{w_0^2}
  \left(
1+
\frac{ik w_0^2}{2z}
  \right)
L_n^{m} (a)
\right) , \nonumber \\
\end{eqnarray}
which yields
\begin{eqnarray}
\hat{l}_{-} \Psi_n^m(r,\phi,z)
=&&
-
2\sqrt{2}
\frac{z}{w_0}
\hbar
\sqrt{n+m}
\Psi_n^{m-1}(r,\phi,z)
\nonumber \\
&& \cdot
\left(
1-
\frac{1}{n+m}
\frac{a}{2}
  \left(
1+
\frac{ik w_0^2}{2z}
  \right)
\frac{L_n^{m} (a)}{L_n^{m-1} (a)}
\right). \nonumber \\
\end{eqnarray}
Therefore, the lowering operator is also successfully working to decrement $m$, independently of the value and the sign of $m$. 

Thus, in this subsection, by obtaining the wavefunction after the ladder operations, we confirmed that the ladder operations work to change the quantised OAM along the propagation direction in units of $\hbar$.

\subsection{Norm after ladder operations}
In this subsection, we obtain the norm of the wavefunctions after ladder operations.
We calculate it for separately depending on the sign of $m$, as in the previous subsection.
We have extensively used the integration formulas, which are summarised in Appendixes.

\subsubsection{Norm of $\hat{l}_{+} \Psi_n^m$ for $m\ge 0$}
First, we rewire the wavefunction by using the formula, $L_n^{m}(a) =L_{n}^{m+1}(a) -L_{n-1}^{m+1}(a)$, as 
\begin{eqnarray}
\hat{l}_{+} 
\Psi_n^m
=&&
-
2\sqrt{2}
\frac{z}{w_0}
\hbar
\sqrt{n+m+1}
\psi_n^{m+1}
\nonumber \\
&&
\left(
\frac{1}{2}
  \left(
1+
\frac{ik w_0^2}{2z}
  \right)
+
\frac{1}{2}
  \left(
1-
\frac{ik w_0^2}{2z}
  \right)
\frac{L_n^{m+1} (a)}{L_n^{m+1} (a)}
\right). \nonumber \\
\end{eqnarray}
Then, we obtain
\begin{eqnarray}
&&
\int_0^{\infty}
2 \pi r dr
|\hat{l}_{+} 
\psi_n^m(r,\phi,z)|^2 \nonumber \\
&&=
2
\left(
\frac{z}{w_0}
\right)^2
\hbar^2
(n+m+1)
  \left(
1-
\frac{ik w_0^2}{2z}
  \right)
  \left(
1+
\frac{ik w_0^2}{2z}
  \right)
\nonumber \\
&&
\cdot
\frac{1}{w_0^2}
\left(
\int_0^{\infty}
2 \pi r dr
\frac{2}{\pi}
\frac{n!}{(n+m+1)!}
a^{m+1}
{\rm e}^{-a}
L_n^{m+1} (a)
L_n^{m+1} (a)
\right.
\nonumber \\
&&
\left.
\ \  +\int_0^{\infty}
2 \pi r dr
\frac{2}{\pi}
\frac{n!}{(n+m+1)!}
a^{m+1}
{\rm e}^{-a}
L_{n-1}^{m+1} (a)
L_{n-1}^{m+1} (a)
\right) \nonumber \\
&&=
\hbar^2
\left(
2
\left(
\frac{z}{w_0}
\right)^2
+
\frac{1}{2}
(k w_0)^2
\right)
(2n+m+1).
\end{eqnarray}

\subsubsection{Norm of $\hat{l}_{+} \Psi_n^m$ for $m< 0$}
Similarly, we obtain 
\begin{eqnarray}
&&
\int_0^{\infty}
2 \pi r dr
|\hat{l}_{+} 
\psi_n^m(r,\phi,z)|^2 \nonumber \\
&=&
8
\left(
\frac{z}{w_0}
\right)^2
\hbar^2
(n-m)
\frac{n!}{(n-m-1)!}
\frac{1}{w_0^2}
w_0^2
\nonumber \\
&&
\cdot
\left(
\int_0^{\infty}
da
{\rm e}^{-a}
a^{-m-1}
L_n^{-m-1} (a)
L_n^{-m-1} (a)
\right.
\nonumber \\
&&
\left.
-
\frac{1}{n-m}
\int_0^{\infty}
da
{\rm e}^{-a}
a^{-m}
L_n^{-m-1} (a)
L_n^{-m} (a)
\right.
\nonumber \\
&&
+
\frac{1}{4 (n-m)^2}
\left(
1+\frac{k^2w_0^4}{4z^2}
\right) \nonumber \\
&&
\left.
\int_0^{\infty}
da
{\rm e}^{-a}
a^{-m+1}
L_n^{-m} (a)
L_n^{-m} (a)
\right)
\nonumber \\
&=&
8
\left(
\frac{z}{w_0}
\right)^2
\hbar^2
(n-m)
\frac{n!}{(n-m-1)!}
\nonumber \\
&&
\cdot
\left(
\frac{(n-m-1)!}{n!}
-
\frac{(n-m)!}{n!}
\frac{1}{n-m}
\right.
\nonumber \\
&&
\left.
+
\frac{1}{4 (n-m)^2}
\left(
1+\frac{k^2w_0^4}{4z^2}
\right)
\frac{(n-m)!}{n!}
(2n-m+1)
\right) \nonumber \\
&=&
\hbar^2
\left(
2
\left(
\frac{z}{w_0}
\right)^2
+
\frac{1}{2}
(k w_0)^2
\right)
(2n-m+1).
\end{eqnarray}

\subsubsection{Norm of $\hat{l}_{-} \Psi_n^m$ for $m\le 0$}
Next, we calculate the norm of $\hat{l}_{-} \Psi_n^m$ for $m\le 0$ as 
\begin{eqnarray}
&&
\int_0^{\infty}
2 \pi r dr
|\hat{l}_{-} 
\psi_n^m(r,\phi,z)|^2 \nonumber \\
&=&
8
\left(
\frac{z}{w_0}
\right)^2
\hbar^2
(n-m+1)
\frac{n!}{(n-m+1)!}
\frac{1}{w_0^2}
w_0^2
\nonumber \\
&&
\cdot
\int_0^{\infty}
da
{\rm e}^{-a}
a^{-m-1} 
\left(
L_n^{-m-1} (a)
L_n^{-m-1} (a)
\right.
\nonumber \\
&&
-
L_n^{-m} (a)
L_n^{-m-1} (a)
+
\frac{1}{4}
\left(
1+\frac{k^2w_0^4}{4z^2}
\right)
L_n^{-m} (a)
L_n^{-m} (a)
) \nonumber \\
&=&
8
\left(
\frac{z}{w_0}
\right)^2
\hbar^2
(n-m+1)
\frac{n!}{(n-m+1)!}
\nonumber \\
&&
\cdot
\left(
\frac{(n-m-1)!}{n!}
-
\frac{(n-m-1)!}{n!}
\right.
\nonumber \\
&&
\left.
+
\frac{1}{4}
\left(
1+\frac{k^2w_0^4}{4z^2}
\right)
\frac{(n-m)!}{n!}
(2n-m+1)
\right) 
\nonumber \\
&&=
\hbar^2
\left(
2
\left(
\frac{z}{w_0}
\right)^2
+
\frac{1}{2}
(k w_0)^2
\right)
(2n-m+1).
\end{eqnarray}

\subsubsection{Norm of $\hat{l}_{-} \Psi_n^m$ for $m> 0$}
Finally, we calculate $\hat{l}_{-} \Psi_n^m$ for $m>0$ as
\begin{eqnarray}
&&
\int_0^{\infty}
2 \pi r dr
|\hat{l}_{-} 
\psi_n^m(r,\phi,z)|^2
\nonumber \\
&=&
8
\left(
\frac{z}{w_0}
\right)^2
\hbar^2
(n+m)
\frac{n!}{(n+m-1)!}
\frac{1}{w_0^2}
w_0^2
\nonumber \\
&&
\cdot
\left(
\int_0^{\infty}
da
{\rm e}^{-a}
a^{m-1}
L_n^{m-1} (a)
L_n^{m-1} (a)
\right.
\nonumber \\
&&
\left.
-
\frac{1}{n+m}
\int_0^{\infty}
da
{\rm e}^{-a}
a^{m}
L_n^{m} (a)
L_n^{m-1} (a)
\right.
\nonumber \\
&&
\left.
+
\frac{1}{4 (n+m)^2}
\left(
1+\frac{k^2w_0^4}{4z^2}
\right)
\int_0^{\infty}
da
{\rm e}^{-a}
a^{m+1}
L_n^{m} (a)
L_n^{m} (a)
\right)
\nonumber \\
&=&
8
\left(
\frac{z}{w_0}
\right)^2
\hbar^2
(n+m)
\frac{n!}{(n+m-1)!}
\nonumber \\
&&
\cdot
\left(
\frac{(n+m-1)!}{n!}
-
\frac{(n+m)!}{n!}
\frac{1}{n+m}
\right.
\nonumber \\
&&
\left.
+
\frac{1}{4 (n+m)^2}
\left(
1+\frac{k^2w_0^4}{4z^2}
\right)
\frac{(n+m)!}{n!}
(2n+m+1)
\right)
\nonumber \\
&=&
\hbar^2
\left(
2
\left(
\frac{z}{w_0}
\right)^2
+
\frac{1}{2}
(k w_0)^2
\right)
(2n+m+1).
\end{eqnarray}

\subsubsection{Summary of the norm of $\hat{l}_{\pm} \Psi_n^m$}
The above direct calculations show that we obtain the same norm for $\hat{l}_{\pm} \Psi_n^m$:
\begin{eqnarray}
&&\int_0^{\infty}
2 \pi r dr
|\hat{l}_{\pm} 
\psi_n^m(r,\phi,z)|^2 \nonumber \\
&&=
\hbar^2
\left(
2
\left(
\frac{z}{w_0}
\right)^2
+
\frac{1}{2}
(k w_0)^2
\right)
(2n+|m|+1),
\end{eqnarray}
which is independent of the sign of $m$.
The first term is coming from the origin dependent extrinsic OAM, while the second term corresponds to the contribution from the intrinsic OAM, which is always positive and finite.
The obtained form of $(\hbar k w_0)^2/2=(p w_0)^2/2$ with the momentum $p=\hbar k$ is intuitively understandable, since $p w_0$ has the dimension of the angular momentum.
As for the case of the plane wave, however, the fact that we obtain $\langle |\hat{l}_{+}\psi_n^m|^2 \rangle=\langle | \hat{l}_{-}\psi_n^m|^2 \rangle$ means that we cannot confirm the validity of the commutation relationship, and thus, we cannot obtain the expectation value of the magnitude of OAM by simple norm calculations.
This might be linked to the fact that the applications of ladder operations contain position dependent factors, such that the orbitals are significantly distorted.
In fact, the LG modes are not eigenstates for the magnitude of the OAM, and they are superposition states with different magnitude of the OAM.
Nevertheless, we can calculate the expectation value of the magnitude of the OAM, as we shall see below.
For that purpose, it is inevitable to calculate $\langle \hat{l}_{+} \hat{l}_{-}\rangle$ and $\langle \hat{l}_{-} \hat{l}_{+}\rangle$, directly, using the LG modes, which are shown in the next subsection.

\subsection{Validity of commutation relationship for LG modes}
First, we must calculate the wavefunction of $\hat{l}_{+} \hat{l}_{-}\Psi_n^m$ and $\hat{l}_{-} \hat{l}_{+}\Psi_n^m$ and calculate the expectation value.
This is straightforward but tedious.
Again, we will split calculations for positive and negative values of $m$.

\subsubsection{Ladder operators}
Before calculating the matrix elements, we will describe operators, $\hat{l}_{+}\hat{l}_{-}$ and $\hat{l}_{-}\hat{l}_{+}$, by using a cylindrical coordinate $(r,\phi,z)$, as 
\begin{eqnarray}
\frac{
  \hat{l}_{+}
  \hat{l}_{-}
  }{\hbar^2}
&=&
\hbar
{\rm e}^{i \phi}
\left(
z \partial_r
+
\frac{z}{r}
i
\partial_{\phi}
-
r 
\partial_z
\right)
\nonumber \\
&&
\hbar
{\rm e}^{-i \phi}
\left(
- z \partial_r
+
\frac{z}{r}
i
\partial_{\phi}
+
r 
\partial_z
\right)
\nonumber \\
&=&
z
\left(
-z\partial_r^2
-
\frac{z^2}{r^2}i \partial_{\phi}
+
\frac{z}{r} i \partial_{r} \partial_{\phi} 
+\partial_z
+ r \partial_r \partial_z
\right)
\nonumber \\
&&+\frac{z}{r}i
\left(
-z\partial_{\phi} \partial_r
+\frac{z}{r}i\partial_{\phi^2}
+
r\partial_{\phi}\partial_z
\right)
\nonumber \\
&&+\frac{z}{r}i(-i)
\left(
-z\partial_{r} 
+\frac{z}{r}i\partial_{\phi}
+
r\partial_z
\right)
\nonumber \\
&&-r
\left(
-\partial_{r} 
-z \partial_z \partial_r
+\frac{i}{r} \partial_{\phi}
+
\frac{z}{r} i \partial_z \partial_{\phi}
+r\partial_z^2
\right)
\nonumber \\
&=&
-z^2 \partial_r^2
-
\frac{z^2}{r}
\partial_r
-
\frac{z^2}{r^2}
\partial_{\phi}^2
\nonumber \\
&&
+(1+2z\partial_z) r \partial_r
-r^2 \partial_z^2
+
2z\partial_z
-i\partial_{\phi}, 
\end{eqnarray}
and 
\begin{eqnarray}
\frac{
  \hat{l}_{-}
  \hat{l}_{+}
  }{\hbar^2}
&=&
\hbar
{\rm e}^{-i \phi}
\left(
- z \partial_r
+
\frac{z}{r}
i
\partial_{\phi}
+
r 
\partial_z
\right)
\hbar
 \nonumber \\
&&
{\rm e}^{i \phi}
\left(
z \partial_r
+
\frac{z}{r}
i
\partial_{\phi}
-
r 
\partial_z
\right)
\nonumber \\
&=&
-z
\left(
z\partial_r^2
-
\frac{z^2}{r^2}i \partial_{\phi}
+
\frac{z}{r} i \partial_{r} \partial_{\phi} 
-\partial_z
- r \partial_r \partial_z
\right)
\nonumber \\
&&+\frac{z}{r}i
\left(
z\partial_{\phi} \partial_r
+\frac{z}{r}i\partial_{\phi^2}
-
r\partial_{\phi}\partial_z
\right)
\nonumber \\
&&
-\frac{z}{r}
\left(
z\partial_{r} 
+\frac{z}{r}i\partial_{\phi}
-
r\partial_z
\right)
\nonumber \\
&&
+r
\left(
\partial_{r} 
+z \partial_z \partial_r
+\frac{i}{r} \partial_{\phi}
+
\frac{z}{r} i \partial_z \partial_{\phi}
-r\partial_z^2
\right)
\nonumber \\
&=&
-z^2 \partial_r^2
-
\frac{z^2}{r}
\partial_r
-
\frac{z^2}{r^2}
\partial_{\phi}^2
\nonumber \\
&&
+(1+2z\partial_z) r \partial_r
-r^2 \partial_z^2
+
2z\partial_z
+i\partial_{\phi}. 
\end{eqnarray}

Thus, we also confirmed the commutation relationship for $\hat{l}_{\pm}$, 
\begin{eqnarray}
\left[
  \hat{l}_{+}
,
  \hat{l}_{-}
\right]
=
2 \hbar  \hat{l}_{z}.
\end{eqnarray}
Therefore, if we apply these operators to a LG mode, we must confirm that this identity is always valid.
This is useful to check the validity of the calculation.

\subsubsection{$\hat{l}_{+} \hat{l}_{-}$ operation for $m\ge0$}
First, we evaluate various terms as follows:
\begin{eqnarray}
\partial_r 
\Psi_n^m
&=&
\frac{1}{w_0}
\sqrt{
\frac{2}{\pi}
\frac{n!}{(n+m)!}
}
\left(
\frac{\sqrt{2}r}{w_0}
\right)^{m}
{\rm  e}^{-\frac{r^2}{w_0^2}}
{\rm  e}^{i m \phi}
{\rm  e}^{i kz}
\nonumber \\
&&
\cdot
\frac{1}{r}
\left[
m
+
2a\frac{d}{da}
-a
\right]
L_n^{m} 
\left(
a \right), 
\end{eqnarray}

\begin{eqnarray}
r
\partial_r 
\Psi_n^m
&&=
\frac{1}{w_0}
\sqrt{
\frac{2}{\pi}
\frac{n!}{(n+m)!}
}
a^{m/2}
{\rm  e}^{-a/2}
{\rm  e}^{i m \phi}
{\rm  e}^{i kz}
\nonumber \\
&&
\cdot
\left[
2(n+m)L_n^{m-1}
-m L_n^{m} 
-aL_n^{m} 
\right],
\end{eqnarray}

\begin{eqnarray}
\frac{1}{r}
\partial_r 
\Psi_n^m
&=&
\frac{1}{w_0}
\sqrt{
\frac{2}{\pi}
\frac{n!}{(n+m)!}
}
a^{m/2}
{\rm  e}^{-a/2}
{\rm  e}^{i m \phi}
{\rm  e}^{i kz}
\nonumber \\
&&
\cdot
\frac{2}{w_0^2}
\left[
-2
L_n^{m+1} 
+ L_n^{m} 
+
\frac{m}{a}
L_n^{m} 
\right],
\end{eqnarray}

\begin{eqnarray}
-z^2
\partial_r^2 
\Psi_n^m
=&&
-2
\frac{z^2}{w_0^2}
\frac{1}{w_0}
\sqrt{
\frac{2}{\pi}
\frac{n!}{(n+m)!}
}
a^{m/2}
{\rm  e}^{-a/2}
{\rm  e}^{i m \phi}
{\rm  e}^{i kz}
\nonumber \\
&&
\cdot
[
m(m-1)\frac{1}{a}
L_n^{m} 
\left(
a
\right) 
+2
L_n^{m+1} 
\left(
a
\right) 
\nonumber \\
 &&
-(4n+2m+3)
L_n^{m} 
\left(
a
\right) 
+aL_n^{m} 
\left(
a
\right)
],
\end{eqnarray}

\begin{eqnarray}
-
\frac{z^2}{r^2}
\partial_{\phi}^2 
\Psi_n^m
=&
2
\frac{z^2}{w_0^2}
m^2
\frac{1}{a}
\Psi_n^m,
\end{eqnarray}

\begin{eqnarray}
-
r^2
\partial_{z}^2 
\Psi_n^m
=&
\frac{1}{2}
k^2 
w_0^2
a
\Psi_n^m,
\end{eqnarray}

\begin{eqnarray}
\left(
r \partial_r
+2z\partial_z r \partial_r
\right)
\Psi_n^m
=&
(1+2ikz)
r \partial_r
\Psi_n^m,
\end{eqnarray}

\begin{eqnarray}
2z\partial_z 
\Psi_n^m
=&
2ikz
\Psi_n^m,
\end{eqnarray}
and finally
\begin{eqnarray}
-i \partial_{\phi}
\Psi_n^m
=&
m\Psi_n^m.
\end{eqnarray}

By summing up some of these terms, we obtain
\begin{eqnarray}
&&\left[
-z^2
\partial_r^2 
-
\frac{z^2}{r}
\partial_r
-
\frac{z^2}{r^2}
\partial_{\phi}^2
\right]
\Psi_n^m \nonumber \\
&&=
2
\frac{z^2}{w_0^2}
\frac{1}{w_0}
\sqrt{
\frac{2}{\pi}
\frac{n!}{(n+m)!}
}
a^{m/2}
{\rm  e}^{-a/2}
{\rm  e}^{i m \phi}
{\rm  e}^{i kz}
\nonumber \\
&&
\cdot
\left[
m(m-1)\frac{1}{a}
L_n^{m} 
\left(
a
\right) 
+2
L_n^{m+1} 
\left(
a
\right) 
\right.
\nonumber \\
&&
\left.
 -(4n+2m+3)
L_n^{m} 
\left(
a
\right) 
+aL_n^{m} 
\left(
a
\right) 
\right.
\nonumber \\
&&
\left.
+2L_n^{m+1} 
\left(
a
\right) 
-
L_n^{m} 
\left(
a
\right) 
-\frac{m}{a}
L_n^{m} 
\left(
a
\right) 
+
\frac{m^2}{a}
L_n^{m} 
\left(
a
\right) 
\right]
\nonumber \\
&=&
2
\frac{z^2}{w_0^2}
\frac{1}{w_0}
\sqrt{
\frac{2}{\pi}
\frac{n!}{(n+m)!}
}
a^{m/2}
{\rm  e}^{-a/2}
{\rm  e}^{i m \phi}
{\rm  e}^{i kz}
\nonumber \\
&&
\cdot
\left[
2(2n+m+1)
L_n^{m} 
\left(
a
\right) 
-
a
L_n^{m} 
\left(
a
\right) 
\right].
\end{eqnarray}

By using the integration formulas (Appendixes), we finally obtain
\begin{eqnarray}
&&\int_0^{\infty}
2 \pi r dr
\left(
\Psi_n^m(r,\phi,z)
\right)^{*}
\hat{l}_{+} (r,\phi,z)
\hat{l}_{-} (r,\phi,z)
\Psi_n^m(r,\phi,z)
\nonumber \\
&&=
\hbar^2
2
\left(
\frac{z}{w_0}
\right)^2
\left(
2(2n+m+1)
-(2n+m+1)
\right)
\nonumber \\
&&
+\hbar^2 (1+2ikz)
\left(
2(n+m)-m-(2n+m+1)
\right)
\nonumber \\
&&+
\hbar^2 \frac{1}{2}(k w_0)^2
(2n+m+1)
+\hbar^2 2ikz
+\hbar^2m
\nonumber \\
&&=
\hbar^2
\left(
2 \left(
\frac{z}{w_0}
\right)^2
+
\frac{1}{2}
(k w_0)^2
\right)
(2n+m+1)
+\hbar^2 (m-1).
\nonumber \\
\end{eqnarray}

\subsubsection{$\hat{l}_{-} \hat{l}_{+}$ operation for $m\ge0$}
The only source of the difference between $\hat{l}_{+} \hat{l}_{-}$ and $\hat{l}_{-} \hat{l}_{+}$ operations is coming from the sign of $i\partial_{\phi}$.
Therefore, it is straightforward to obtain
\begin{eqnarray}
&\int_0^{\infty}
2 \pi r dr
\left(
\Psi_n^m(r,\phi,z)
\right)^{*}
\hat{l}_{-} (r,\phi,z)
\hat{l}_{+} (r,\phi,z)
\Psi_n^m(r,\phi,z)
\nonumber \\
&=
\hbar^2
\left(
2 \left(
\frac{z}{w_0}
\right)^2
+
\frac{1}{2}
(k w_0)^2
\right)
(2n+m+1)
-\hbar^2 (m+1)
\nonumber \\
\end{eqnarray}

This result, together with the previous result for $\langle  \hat{l}_{-} \hat{l}_{+} \rangle$, confirms that the commutation relationship over average indeed satisfies 
\begin{eqnarray}
\left \langle
\left[
  \hat{l}_{+}
,
  \hat{l}_{-}
\right]
\right \rangle
=
2 \hbar  
\left \langle
 \hat{l}_{z}
\right \rangle
,
\end{eqnarray}
where $\langle  \hat{l}_{z} \rangle = \hbar m$ for $m\ge0$.

\subsubsection{$\hat{l}_{+} \hat{l}_{-}$ operation for $m\le0$}
We can proceed for $m\le0$ in a similar way.
First, we evaluate the factors:
\begin{eqnarray}
r
\partial_r
\Psi_n^m
=&&
\frac{1}{w_0}
\sqrt{
\frac{2}{\pi}
\frac{n!}{(n-m)!}
}
a^{-m/2}
{\rm  e}^{-a/2}
{\rm  e}^{i m \phi}
{\rm  e}^{i kz}
\nonumber \\
&&\cdot
\left[
2(n-m)L_n^{-m-1} (a)
+mL_n^{-m} (a)
-aL_n^{-m} (a)
\right], \nonumber \\
\end{eqnarray}

\begin{eqnarray}
\frac{1}{r}
\partial_r
\Psi_n^m
=&&
\frac{1}{w_0}
\sqrt{
\frac{2}{\pi}
\frac{n!}{(n-m)!}
}
a^{-m/2}
{\rm  e}^{-a/2}
{\rm  e}^{i m \phi}
{\rm  e}^{i kz}
\nonumber \\
&& \cdot
\frac{2}{w_0^2}
\left[
-2 L_n^{-m+1} (a)
+L_n^{-m} (a)
-\frac{m}{a} L_n^{-m} (a)
\right], \nonumber \\
\end{eqnarray}

\begin{eqnarray}
-z^2
\partial_r^2 
\Psi_n^m
=&&
-2
\frac{z^2}{w_0^2}
\frac{1}{w_0}
\sqrt{
\frac{2}{\pi}
\frac{n!}{(n-m)!}
}
a^{-m/2}
{\rm  e}^{-a/2}
{\rm  e}^{i m \phi}
{\rm  e}^{i kz}
\nonumber \\
&&
\cdot
[
m(m+1)\frac{1}{a}
L_n^{-m} 
\left(
a
\right) 
+2
L_n^{-m+1} 
\left(
a
\right) 
\nonumber \\
&&-(4n-2m+3)
L_n^{-m} 
\left(
a
\right) 
+aL_n^{-m} 
\left(
a
\right) 
],
\end{eqnarray}
and the other factors are the same ones for $m\ge0$.
Then, we obtain 
\begin{eqnarray}
&&\left[
-z^2
\partial_r^2 
-
\frac{z^2}{r}
\partial_r
-
\frac{z^2}{r^2}
\partial_{\phi}^2
\right]
\Psi_n^m
\nonumber \\
&=&
2
\frac{z^2}{w_0^2}
\frac{1}{w_0}
\sqrt{
\frac{2}{\pi}
\frac{n!}{(n-m)!}
}
a^{-m/2}
{\rm  e}^{-a/2}
{\rm  e}^{i m \phi}
{\rm  e}^{i kz}
\nonumber \\
&&
\cdot
\left[
-2L_n^{-m+1}(a)
+(4n-2m+3)L_n^{-m}(a)
\right.
\nonumber \\
&&
\left.
-m(m+1)\frac{1}{a}
L_n^{-m} 
\left(
a
\right) 
-a L_n^{-m} 
\right.
\nonumber \\
&&
\left.
2L_n^{-m+1} 
\left(
a
\right) 
-L_n^{-m} 
\left(
a
\right) 
\right.
\nonumber \\
&&
\left.
+\frac{m}{a}L_n^{-m} 
\left(
a
\right) 
+\frac{m^2}{a}
L_n^{-m} 
\left(
a
\right) 
\right]
\nonumber \\
&=&
2
\frac{z^2}{w_0^2}
\frac{1}{w_0}
\sqrt{
\frac{2}{\pi}
\frac{n!}{(n-m)!}
}
a^{-m/2}
{\rm  e}^{-a/2}
{\rm  e}^{i m \phi}
{\rm  e}^{i kz}
\nonumber \\
&&
\cdot
\left[
2(2n+m+1)
L_n^{-m} 
\left(
a
\right) 
-
a
L_n^{-m} 
\left(
a
\right) 
\right].
\end{eqnarray}
Therefore, we finally obtain
\begin{eqnarray}
&&
\int_0^{\infty}
2 \pi r dr
\left(
\Psi_n^m(r,\phi,z)
\right)^{*}
\hat{l}_{+} (r,\phi,z)
\hat{l}_{-} (r,\phi,z)
\Psi_n^m(r,\phi,z)
\nonumber \\
&&=
\hbar^2
2
\left(
\frac{z}{w_0}
\right)^2
\left(
2(2n-m+1)
-(2n-m+1)
\right)
\nonumber \\
&&
+\hbar^2 (1+2ikz)
\left(
2(n-m)+m-(2n-m+1)
\right)
\nonumber \\
&& +
\hbar^2 \frac{1}{2}(k w_0)^2
(2n-m+1)
+\hbar^2 2ikz
+\hbar^2m
\nonumber \\
&&=
\hbar^2
\left(
2 \left(
\frac{z}{w_0}
\right)^2
+
\frac{1}{2}
(k w_0)^2
\right)
(2n-m+1)
+\hbar^2 (m-1).
\nonumber \\
\end{eqnarray}

\subsubsection{$\hat{l}_{-} \hat{l}_{+}$ operation for $m\le0$}
Again, the only source of the change between $\hat{l}_{+} \hat{l}_{-}$ and $\hat{l}_{-} \hat{l}_{+}$ operations is coming from the sign of $i\partial_{\phi}$, such that we obtain

\begin{eqnarray}
&&\int_0^{\infty}
2 \pi r dr
\left(
\Psi_n^m(r,\phi,z)
\right)^{*}
\hat{l}_{-} (r,\phi,z)
\hat{l}_{+} (r,\phi,z)
\Psi_n^m(r,\phi,z)
\nonumber \\
&&=
 \hbar^2
\left(
2\left(
\frac{z}{w_0}
\right)^2
+
\frac{1}{2}
(k w_0)^2
\right)
(2n-m+1)
-\hbar^2 (m+1).
\nonumber \\
\end{eqnarray}
Therefore, the commutation relationship over average
\begin{eqnarray}
\left \langle
\left[
  \hat{l}_{+}
,
  \hat{l}_{-}
\right]
\right \rangle
=
2 \hbar  
\left \langle
 \hat{l}_{z}
\right \rangle
,
\end{eqnarray}
is also valid for $m\le0$.

\subsection{Summary of expectation values of $\hat{l}_{+} \hat{l}_{-}$ and $\hat{l}_{-} \hat{l}_{+}$ operations}
The above results are summarised as follows,
\begin{eqnarray}
&&\int_0^{\infty}
2 \pi r dr
\left(
\Psi_n^m(r,\phi,z)
\right)^{*}
\hat{l}_{+} (r,\phi,z)
\hat{l}_{-} (r,\phi,z)
\Psi_n^m(r,\phi,z)
\nonumber \\
&&=
 \hbar^2
\left(
2\left(
\frac{z}{w_0}
\right)^2
+
\frac{1}{2}
(k w_0)^2
\right)
(2n+|m|+1)
+\hbar^2 (m-1) ,
\nonumber \\
\end{eqnarray}
and 
\begin{eqnarray}
&&\int_0^{\infty}
2 \pi r dr
\left(
\Psi_n^m(r,\phi,z)
\right)^{*}
\hat{l}_{-} (r,\phi,z)
\hat{l}_{+} (r,\phi,z)
\Psi_n^m(r,\phi,z)
\nonumber \\
&&=
 \hbar^2
\left(
2\left(
\frac{z}{w_0}
\right)^2
+
\frac{1}{2}
(k w_0)^2
\right)
(2n+|m|+1)
-\hbar^2 (m+1) ,
\nonumber \\
\end{eqnarray}
which are independent of the sign of $m$.
The commutation relationship over average
\begin{eqnarray}
\left \langle
\left[
  \hat{l}_{+}
,
  \hat{l}_{-}
\right]
\right \rangle
=
2 \hbar  
\left \langle
 \hat{l}_{z}
\right \rangle
,
\end{eqnarray}
is also valid for $^{\forall}m$.

\subsection{Magnitude of OAM}
The above calculations have confirmed the quantum commutation relationship, $[\hat{l}_{+},\hat{l}_{-}
]=2 \hbar  \hat{l}_{z}$, as an expectation value after the application to the LG mode.
This is trivial, because the commutation relationship is valid at the operator level, such that it should be valid even after the application to the LG mode.
Nevertheless, obtained matrix elements of expectation values of $\hat{l}_{+}\hat{l}_{-}$ and $\hat{l}_{-}\hat{l}_{+}$ are useful to evaluate the magnitude of the OAM, because the inner product of the vectorial OAM operator is described as
\begin{eqnarray}
\hat{\bf l} \cdot \hat{\bf l}
&=&
\hat{l}_x^2
+
\hat{l}_y^2
+
\hat{l}_z^2 \\
&=&
\hat{l}_+
\hat{l}_-
+
\hat{l}_z^2
-
\hbar 
\hat{l}_z \\
&=&
\hat{l}_-
\hat{l}_+
+
\hat{l}_z^2
+
\hbar 
\hat{l}_z.
\end{eqnarray}
We can confirm that the expectation value is independent of whether we are using  $\langle \hat{l}_{+}\hat{l}_{-}\rangle$ or $\langle \hat{l}_{-}\hat{l}_{+}\rangle$, and we obtain
\begin{eqnarray}
&&
\int_0^{\infty}
2 \pi r dr
\Psi_n^{m}(r,\phi,z)^{*}
\hat{\bf l} \cdot \hat{\bf l}
\Psi_n^m (r,\phi,z) \nonumber \\
&=&
\int_0^{\infty}
2 \pi r dr
\Psi_n^{m}(r,\phi,z)^{*}
\hat{l}_{+} 
\hat{l}_{-}
\Psi_n^m (r,\phi,z) 
+\hbar^2 m (m-1)
\nonumber \\
&=&
\int_0^{\infty}
2 \pi r dr
\Psi_n^{m}(r,\phi,z)^{*}
\hat{l}_{-} 
\hat{l}_{+}
\Psi_n^m (r,\phi,z) 
+\hbar^2 m (m+1)
\nonumber \\
&=&
\hbar^2
\left(
2 \left(
\frac{z}{w_0}
\right)^2
+
\frac{1}{2}
(k w_0)^2
\right)
(2n+|m|+1) \nonumber \\
&&+\hbar^2 (m+1)(m-1).
\end{eqnarray}

The expectation value does not depend on the sign of $m$, which is reasonable in a system with a helical symmetry.
The first term contains the origin ($z=0$) dependent contribution of the extrinsic OAM.
If we take the average over $z\in(0,L)$, we obtain
\begin{eqnarray}
\langle \hat{\bf l} \cdot \hat{\bf l} \rangle
&=&
\hbar^2
\left(
\frac{2}{3} 
\left(
\frac{L}{w_0}
\right)^2
+
\frac{1}{2}
(k w_0)^2
\right)
(2n+|m|+1) \nonumber \\
&&+\hbar^2 (m+1)(m-1),
\end{eqnarray}
while if we average over  $z\in(-L/2,L/2)$, it becomes
\begin{eqnarray}
\langle \hat{\bf l} \cdot \hat{\bf l} \rangle
&=&
\hbar^2
\left(
\frac{1}{6} 
\left(
\frac{L}{w_0}
\right)^2
+
\frac{1}{2}
(k w_0)^2
\right)
(2n+|m|+1) \nonumber \\
&&+\hbar^2 (m+1)(m-1).
\end{eqnarray}

The other contributions are from intrinsic OAM.
If $m\gg1$, the most of the energy of photons is used to sustain the rotating motion as OAM, such that $\hbar \delta \omega_0 m \gg \hbar v_0 k $ in the dispersion relationship. 
In this limit, the intrinsic OAM is dominated by the contribution from $\hbar m$, which is consistent with the above formula.

In the opposite limit of the absence of the definite quantised OAM ($n=m=0$), the beam becomes a simple Gaussian wave.
Even in this case, 
\begin{eqnarray}
\left \langle
\hat{l}_{\rm intrinsic}^2
\right \rangle
\rightarrow
\frac{1}{2}
(\hbar k w_0)^2
-\hbar^2 
\sim 
\frac{1}{2}
(p w_0)^2
-\hbar^2
\end{eqnarray}
holds, which is an intuitive formula, because the same amount of the angular momentum magnitude with spin of $\hbar$ is subtracted.
The finite intrinsic OAM is coming from quantum mechanical fluctuations due to $\hat{l}_x$, $\hat{l}_y$, and quantum commutation relationship.
Even if the quantum number of the mode is zero ($m=0$), the quantum fluctuation is inevitable, such that the finite value of the magnitude remains as the zero-point fluctuation.
If the spin component of $\hbar$ is negligible, the OAM fluctuation is of the order of
\begin{eqnarray}
\sqrt{\left \langle
\hat{l}^2
\right \rangle}
\sim 
p \frac{w_0}{\sqrt{2}}
=
p w_1,
\end{eqnarray}
where $w_1=w_0/\sqrt{2}$ is the effective waist for the OAM.
The total amount of fluctuation as a coherent laser beam is obtained by multiplying the number of photons, $N$.

\subsection{Transfer matrix element}
Finally, we calculate the matrix elements of $\hat{l}_{\pm}$ among the LG modes with different $m$.
In order to allow these to couple, the phase matching condition must be satisfied, such that the energy, $\hbar \omega$, and the momentum, $p=\hbar k$, would be conserved through the operation of $\hat{l}_{\pm}$.
Otherwise, the transfer matrix elements would vanish due to the destructive interference upon propagation.
Strictly speaking, such a condition would not be satisfied during the propagation in the GRIN fibre, since the value of $k$ would be different among modes with different $m$ due to the dispersion relationship.
Therefore, the coupling is expected only in the limit of $g\rightarrow 0$  as a Gaussian beam, where the dispersion is almost negligible and the material is considered to be almost uniform.
We also assume that the beam is sufficiently collimated, such that the impact of the Gouy phase is negligible.
We then obtain 
\begin{eqnarray}
&&
\int_{0}^{\infty}
dr 2 \pi r
\left(
\psi_n^{m+1}(r,\phi,z)
\right)^{*}
\hat{l}_{+} 
\psi_n^m(r,\phi,z) \nonumber \\
&&=
-
\hbar
\frac{z}{w_0}
\sqrt{2(n+m+1)}
\left(
1
+
i
\frac{k w_0^2}{2z}
\right)
\end{eqnarray}
for $m\ge 0$,
\begin{eqnarray}
&&
\int_{0}^{\infty}
dr 2 \pi r
\left(
\psi_n^{m+1}(r,\phi,z)
\right)^{*}
\hat{l}_{+} 
\psi_n^m(r,\phi,z)  
\nonumber \\
&&=
\hbar
\frac{z}{w_0}
\sqrt{2(n+|m|)}
\left(
1
-
i
\frac{k w_0^2}{2z}
\right)
\end{eqnarray}
for $m<0$,
\begin{eqnarray}
&&\int_{0}^{\infty}
dr 2 \pi r
\left(
\psi_n^{m-1}(r,\phi,z)
\right)^{*}
\hat{l}_{-} 
\psi_n^m(r,\phi,z)
\nonumber \\
&&=
-
\hbar
\frac{z}{w_0}
\sqrt{2(n+m)}
\left(
1
-
i
\frac{k w_0^2}{2z}
\right)
\end{eqnarray}
for $m>0$, and 
\begin{eqnarray}
&&
\int_{0}^{\infty}
dr 2 \pi r
\left(
\psi_n^{m-1}(r,\phi,z)
\right)^{*}
\hat{l}_{-} 
\psi_n^m(r,\phi,z)
\nonumber \\
&&=
\hbar
\frac{z}{w_0}
\sqrt{2(n+|m|+1)}
\left(
1
+
i
\frac{k w_0^2}{2z}
\right)
\end{eqnarray}
for $m\le 0$, respectively,
If we take the average over $z\in (0,T)$, assuming a thickness of $T$ for the optical plate to increment or decrement the value of $m$, these results show 
\begin{eqnarray}
\left(
\left \langle
m+1
\right |
\hat{l}_{+}
\left |
m
\right \rangle
\right)^{*}
&=&
-
\hbar
\sqrt{2(n+m+1)}
\frac{1}{2}
\left(
\frac{T}{w_0}
-
i
k w_0
\right) \nonumber \\
&=&
\left \langle
m
\right |
\hat{l}_{-}
\left |
m+1
\right \rangle
\end{eqnarray}
for $m\ge 0$, and 
\begin{eqnarray}
\left(
\left \langle
m+1
\right |
\hat{l}_{+}
\left |
m
\right \rangle
\right)^{*} 
&=&
-
\hbar
\sqrt{2(n+|m|)}
\frac{1}{2}
\left(
\frac{T}{w_0}
+
i
k w_0
\right) \nonumber \\
&=&
\left \langle
m
\right |
\hat{l}_{-}
\left |
m+1
\right \rangle
\end{eqnarray}
for $m<0$.
For both cases, the relationship, 
\begin{eqnarray}
\left(
\left \langle
m+1
\right |
\hat{l}_{+}
\left |
m
\right \rangle
\right)^{*} 
&=&
\left \langle
m
\right |
\hat{l}_{-}
\left |
m+1
\right \rangle,
\end{eqnarray}
is always valid for $^{\forall}m$.
This implies 
\begin{eqnarray}
\hat{l}_{+}
^{\dagger}
&=
\hat{l}_{-}
\\
\hat{l}_{-}
^{\dagger}
&=
\hat{l}_{+}
\end{eqnarray}
for a system described at least in a Hilbert space spanned by the LG modes.
These results also suggest that OAM can be a proper quantum mechanical observable, satisfying the commutation relationship, at least for a system described by the LG modes.
Experimentally, there are many successful demonstrations to control $m$ \cite{Marrucci06,Machavariani07,Lai08,Guan13,Sun14,Naido16,Dorney19}.
In this paper, we have provided a theoretical justification for enabling the increment or decrement of the quantum number, $m$, thus confirming quantum-mechanical description of the OAM.

\section{Conclusions}
A photon, an elementary particle with the internal spin degree of freedom, can have an orbital degree of freedom \cite{Allen92,Enk94,Barnett16,Yariv97,Jackson99,Grynberg10,Bliokh15}.
In a vacuum, a photon travels at the speed of light, $c$, and is described by a plane wave \cite{Jackson99}.  The many-body state for photons can be described by a QED theory \cite{Dirac30,Sakurai67,Fox06,Grynberg10,Bliokh15}.
On the other hand, for a photon confined to a region with a larger refractive index, i.e., a waveguide, the mode is described as a confined mode, which is a bound state with a discrete energy level.
The nature of this mode is completely different from that described by a plane wave allowing a continuous energy spectrum.
We have shown that the fundamental equation to describe the orbital wavefunction of photons in a waveguide is Helmholtz equation \cite{Yariv97,Jackson99} for a monochromatic coherent ray emitted from a laser, where the spin degree of freedom is described by a Jones vector.
We must solve the Helmholtz equation in a material including the refractive index profile and the symmetry of the system.
The reason why the many-body photonic state can be described by a single wavefunction is based on the Bose-Einstein condensation nature of a coherent state that allows a macroscopic number of photons to occupy the lowest energy state, because of the Bose statistics due to the integer spin.
As a specific example, we have considered a GRIN fibre with a cylindrical symmetry, for which we could solve the Helmholtz equation exactly by using LG modes \cite{Allen92,Enk94,Barnett16,Yariv97,Jackson99,Grynberg10,Bliokh15}.

We have defined canonical OAM operators in a cylindrical coordinate and have applied them to the LG modes.
We have confirmed that the OAM is quantised along the direction of propagation and that the quantum-mechanical expectation value is indeed obtained as $\hbar m$, while the average values along the directions perpendicular to the propagation vanish.
We have found that the ladder operators to increase or decrease $m$ work successfully to increment or decrement in units of $\hbar$.
We could also calculate the quantum-mechanical average of the magnitude of OAM as a function of the radial quantum numbers of $n$ and $m$.
We have also confirmed the contributions from the intrinsic OAM and the origin-dependent extrinsic OAM.
Finally, we have calculated the matrix elements of the ladder operators and have confirmed that the angular momentum operators are observable at least in the Hilbert space spanned by the LG modes.
From those results, we conclude that the OAM is a proper quantum-mechanical degree of freedom and that a standard quantum-mechanical treatment is applicable to a monochromatic coherent ray of photons.

\section*{Acknowledgements}
This work is supported by JSPS KAKENHI Grant Number JP 18K19958.
The author would like to express sincere thanks to Prof I. Tomita for continuous discussions and encouragements. 

\appendix

\section{Laguerre function}
We describe full details of Laguerre and associate Laguerre functions and related formulas in Appendixes \cite{Arfken05,Whittaker62,Bateman53}.
First,  we consider a differential equation
\begin{eqnarray}
\left[
a
\frac{d^2}{da^2}
+
(1-a)
\frac{d}{da}
+n
\right]
f
=0,
\end{eqnarray}
which will be solved by assuming a Taylor series expansion
\begin{eqnarray}
f=
\sum_{j=0}^{\infty}
a_j a^j,
\end{eqnarray}
which gives
\begin{eqnarray}
\sum_{j=0}^{\infty}
a^j
\left[
(j+1)^2
a_{j+1}
+
(n-j)
a_j
\right]
=0.
\end{eqnarray}
This provides a recurrence formula
\begin{eqnarray}
a_{j+1}
=
-
\frac{(n-j)^2}{(j+1)^2}
a_j
\end{eqnarray}
for $j=0, 1, 2, \cdots, n$, and $a_{j}=0$ for $j>n$. 
Therefore, we obtain
\begin{eqnarray}
a_{j}
=
(-1)^j
\frac{n !}{(j !)^2 (n-j) !}
a_0.
\end{eqnarray}
The differential equation cannot be determined uniquely without providing a boundary condition.
The same is true for a special function, such that there exists a room to choose the arbitrary value of $a_0$, while it is a standard rule to choose $a_0>0$ in mathematics.
Our definition in this paper is $a_0=1$, but other people are also using $a_0=n!$ as an alternative definition.
In this way, we obtain the solution, $f=f_n(a)$, as 
\begin{eqnarray}
L_{n}(a)
&=&
\sum_{j=0}^{n}
\frac{(-1)^j}{j !}
\frac{n !}{(j !) (n-j) !}
a^j \\
&=&
\sum_{j=0}^{n}
\frac{(-1)^j}{j !}
\ _nC_j
\ a^j,
\end{eqnarray}
where$\ _nC_j$ is a binomial coefficient.

\subsection{Rising operator}
By directly calculating the derivative, we obtain
\begin{eqnarray}
\left [ 
a
\frac{d}{da}
-a
\right ]
L_{n}(a)
&=
\sum_{j=0}^{n+1}
\frac{(-1)^j}{j !}
\frac{(n+1) !}{j ! (n+1-j) !}
j
a^j,
\end{eqnarray}
which can be combined with this identity
\begin{eqnarray}
(n+1)
L_{n}(a)
&=
\sum_{j=0}^{n+1}
\frac{(-1)^j}{j !}
\frac{(n+1) !}{j ! (n+1-j) !}
(n+1-j)
a^j, \nonumber \\
\end{eqnarray}
to obtain
\begin{eqnarray}
\left [ 
a
\frac{d}{da}
-a
+(n+1)
\right ]
L_{n}(a)
=
(n+1)
L_{n+1}(a). \nonumber \\
\end{eqnarray}
This formula works as a raising operator to increase the radial index, $n$.

\subsection{Lowering operator}
Quite similarly, we can also obtain the lowering operator.
Calculating a derivative, 
\begin{eqnarray}
a
\frac{d}{da}
L_{n}(a)
&=&
\sum_{j=0}^{n}
\frac{(-1)^j}{j !}
\frac{n !}{j ! (n-j) !}
j
a^j
\\
&=&
\sum_{j=0}^{n-1}
\frac{(-1)^j}{j !}
\frac{(n-1) !}{j ! (n-1-j) !}
\frac{n}{(n-j)}
j
a^j \nonumber \\
&&+
n
\frac{(-1)^n}{n!}a^n \nonumber \\
\end{eqnarray}
together with the identity
\begin{eqnarray}
-n
L_{n}(a)
&=
\sum_{j=0}^{n-1}
\frac{(-1)^{j}}{j !}
\frac{n !}{j ! (n-1-j) !}
(-n)
\frac{n}{n-j}
a^j
-\frac{(-1)^n}{n!}
a^n, \nonumber \\
\end{eqnarray}
we obtain the lowering operation formula, 
\begin{eqnarray}
\left [ 
a
\frac{d}{da}
-n
\right ]
L_{n}(a)
&=
-n
L_{n-1}(a).
\end{eqnarray}

By summing up raising and lowering operators, we also obtain the recurrence relationship
\begin{eqnarray}
(2n+1-a)
L_{n}(a)
&=
(n+1)
L_{n+1}(a)
+n
L_{n-1}(a), \nonumber \\
\end{eqnarray}
which correlate 3 successive Laguerre functions.

\subsection{Generating function}
The generating function is defined as a function, whose coefficients of series expansion are  Laguerre functions. 
Therefore, it is defined as 
\begin{eqnarray}
G(t,\tau)
=\sum_{i=0}^{\infty}
L_{i}(t)
\tau^i.
\end{eqnarray}
By inserting the series expansion form  of $L_{i}(t)$, we obtain 
\begin{eqnarray}
G(t,\tau)
&=&
\sum_{i=0}^{\infty}
\sum_{j=0}^{i}
\frac{(-1)^j}{j !}
\frac{i !}{j ! (i-j) !}
t^j
\tau^i \\
&=&
\sum_{i=0}^{\infty}
\sum_{k=0}^{\infty}
\frac{(-1)^j}{j !}
\frac{(k+j) !}{j ! k !}
t^j
\tau^{k+j}, 
\end{eqnarray}
where we used $k=i-j$ in the 2nd line.
Together with the binomial theorem
\begin{eqnarray}
\left(
\frac{1}{1-\tau}
\right)
^{j+1}
=
\sum_{k=0}^{\infty}
\frac{(k+j)!}{k! j!}
\tau^k, 
\end{eqnarray}
we finally obtain the analytic formula for the generating function as 
\begin{eqnarray}
G(t,\tau)
&=
\frac{1}{1-\tau}
\sum_{j=0}^{\infty}
\frac{1}{j !}
\left(
-
\frac{t \tau}{1-\tau}
\right)
^{j}
\\
&=
\frac{1}{1-\tau}
\exp
\left(
-
\frac{t \tau}{1-\tau}
\right).
\end{eqnarray}

Using this generating function, we can obtain the orthogonality relationship, which is used to confirm the orthogonality against modes with different radial numbers and calculate the normalisation factors.
In order to derive it, we evaluate the following sum of the integrals, 
\begin{eqnarray}
& &
\sum_{i=0}^{\infty}
\sum_{i'=0}^{\infty}
\tau^i
\tau'^{i'}
\int_{0}^{\infty}
dt
{\rm e}^{-t}
L_{i}(t)
L_{i'}(t)
\nonumber \\
&=&
\frac{1}{(1-\tau)(1-\tau')}
\int_{0}^{\infty}
dt
\exp
\left(
-t
\frac{1-\tau \tau'}{(1-\tau)(1-\tau')}
\right) \nonumber \\
&=&
\frac{1}{1-\tau\tau'}
\left [\exp
\left(
-t
\frac{1-\tau \tau'}{(1-\tau)(1-\tau')}
\right)
\right]_{0}^{\infty}
\nonumber \\
&=&
\frac{1}{1-\tau\tau'}\nonumber  \\
&=&
\sum_{i}^{\infty} (\tau\tau')^i \nonumber \\
&=&
\sum_{i}^{\infty} 
\sum_{i'}^{\infty} 
\delta_{i,i'}
\tau^i \tau'^{i'}.
\end{eqnarray}
Comparing the first term and the last one, we obtain
\begin{eqnarray}
\int_{0}^{\infty}
dt
\ 
{\rm e}^{-t}
L_{i}(t)
L_{i'}(t)
=
\delta_{i,i'}.
\end{eqnarray}

\subsection{Rodrigues formula}
Rodrigues formula is an operator form of the representation of Laguerre function, which will be suitable for quantum mechanics.
In order to obtain it, we just need to evaluate the following function
\begin{eqnarray}
{\rm e}^{t}
\frac{d^n}{dt^n}
\left(
t^n
{\rm e}^{-t}
\right)
&=&
\sum_{j=0}^{n}
\frac{n!}{j!(n-j!)}
\frac{n!}{(n-j!)}
(-1)^{n-j}
t^{n-j} \nonumber \\
&=&
n!
\sum_{j=0}^{n}
\frac{(-1)^{j}}{j!}
\frac{n!}{j!(n-j!)}
t^{j} \nonumber \\
&=& n! L_n(t),
\end{eqnarray}
and thus, we obtain 
\begin{eqnarray}
L_n(t)
=
\frac{1}{n!} 
{\rm e}^{t}
\frac{d^n}{dt^n}
\left(
t^n
{\rm e}^{-t}
\right).
\end{eqnarray}

\section{Associated Laguerre function}
The associated Laguerre function is defined as
\begin{eqnarray}
L_n^m(t)=
(-1)^m
\frac{d^m}{dt^m} 
L_{n+m}(t).
\end{eqnarray}
The factor  of $(-1)^m$ guarantees the first Taylor series expansion coefficient of $a_0$ to be positive $a_0>0$ as a mathematical convention.

The differential equation for the associated Laguerre function is derived from that of the Laguerre function
\begin{eqnarray}
\left [ 
t
\frac{d^2}{dt^2}
+(1-t)
\frac{d}{dt}
+n
\right ]
L_{n}(t)
=
0,
\end{eqnarray}
by the $m$-th derivative of this equation,
\begin{eqnarray}
\left [
\frac{d^m}{dt^m} 
\left(
t
\frac{d^2}{dt^2}
\right)
+
\frac{d^m}{dt^m} 
\left(
(1-t)
\frac{d}{dt}
\right)
+n
\frac{d^m}{dt^m} 
\right ]
L_{n}(t)
=
0, \nonumber \\
\end{eqnarray}
which becomes, 
\begin{eqnarray}
\left [
t
\frac{d^2}{dt^2}
+
(m+1-t)
\frac{d}{dt}
+(n-m)
\right ]
(-1)^m
\frac{d^m}{dt^m} 
L_{n}(t)
=
0. \nonumber \\
\end{eqnarray}
By exchanging $n\rightarrow n+m$, we obtain
\begin{eqnarray}
\left [
t
\frac{d^2}{dt^2}
+
(m+1-t)
\frac{d}{dt}
+n
\right ]
L_{n}^{m}(t)
=
0.
\end{eqnarray}

We also obtain the Taylor series expansion of $L_{n}^{m}(t)$ by direct calculation.
Inserting the series expression for $L_{n}(t)$ into the definition, we obtain
\begin{eqnarray}
L_n^m(t)
&=&
(-1)^m
\frac{d^m}{dt^m} 
\sum_{j=0}^{n+m}
\frac{(-1)^j}{j !}
\frac{(n+m) !}{j ! (n+m-j) !}
t^j \\
&=&
(-1)^m
\sum_{j=m}^{n+m}
\frac{(-1)^j}{j !}
\frac{(n+m) !}{j ! (n+m-j) !}
\frac{j!}{(j-m)!}t^{j-m}
\nonumber \\
&=&
\sum_{j=0}^{n}
\frac{(-1)^j}{j !}
\frac{(n+m) !}{(j+m)! (n-j) !}
t^{j},
\end{eqnarray}
which shows that the term at $j=0$ is indeed positive.

\subsection{Generating function}
The generating function for the associated Laguerre function is defined by
\begin{eqnarray}
G(t,\tau)
&=&\sum_{i=0}^{\infty}
L_{i}^{m} (t)
\tau^i \\
&=&\sum_{i=0}^{\infty}
\sum_{j=0}^{n}
\frac{(-1)^j}{j !}
\frac{(n+m) !}{(j+m)! (n-j) !}
t^{j}
\tau^i \nonumber \\
&=&\sum_{j=0}^{\infty}
\sum_{k=0}^{\infty}
\frac{(-1)^j}{j !}
\frac{(j+k+m) !}{(j+m)! k !}
t^{j}
\tau^{j+k}.
\end{eqnarray}
Using the binomial theorem,
\begin{eqnarray}
\left(
\frac{1}{1-\tau}
\right)
^{j+m+1}
=
\sum_{k=0}^{\infty}
\frac{(k+j+m)!}{k! (j+m)!}
x^k,
\end{eqnarray}
we obtain 
\begin{eqnarray}
G(t,\tau)
&=&\sum_{j=0}^{\infty}
\frac{(-1)^j}{j !}
\left(
\frac{1}{1-\tau}
\right)
^{j+m+1}
\left(
t \tau
\right)
^{j}
\\
&=&
\left(
\frac{1}{1-\tau}
\right)
^{m+1}
\sum_{j=0}^{\infty}
\frac{(-1)^j}{j !}
\left(
-
\frac{t \tau}{1-\tau}
\right)
^{j}
\\
&
=&
\left(
\frac{1}{1-\tau}
\right)
^{m+1}
\exp
\left(
-
\frac{t \tau}{1-\tau}
\right).
\end{eqnarray}

\subsection{Recurrence relationship}
We obtain the recurrence relationship for the associated Laguerre function.
By calculating the derivative of the generating function by $\tau$, we obtain
\begin{eqnarray}
&& \sum_{n=0}^{\infty}
L_{n}^{m} (t)
n
\tau^{n-1} \nonumber \\
&=&
(m+1)
\frac{{\rm e}^{-\frac{t \tau}{1-\tau}} }
  {(1-\tau)^{m+2}}
-
\frac{{\rm e}^{-\frac{t \tau}{1-\tau}} }
  {(1-\tau)^{m+1}}
\frac{t(1-\tau)+t \tau}{(1-\tau)^2}
\nonumber \\
&=&
\frac{m+1}{1-\tau}
\sum_{n=0}^{\infty}
L_{n}^{m} (t)
\tau^{n}
-
\frac{t}{(1-\tau)^2}
\sum_{n=0}^{\infty}
L_{n}^{m} (t)
\tau^{n}. 
\end{eqnarray}
Then, we obtain
\begin{eqnarray}
&&
\sum_{n=0}^{\infty}
L_{n}^{m} (t)
n
\tau^{n-1}
(1-\tau)^2 \nonumber \\
&=&
\frac{m+1}{1-\tau}
\sum_{n=0}^{\infty}
L_{n}^{m} (t)
\tau^{n}
(1-\tau)
-
t
\sum_{n=0}^{\infty}
L_{n}^{m} (t)
\tau^{n}, \nonumber \\
\end{eqnarray}
from which we obtain the recurrence relationship
\begin{eqnarray}
&& (n+1)
L_{n+1}^{m} (t)
-
(2 n +m+1-t) 
L_{n}^{m} (t) \nonumber \\
&& \ \ 
+
(n+m)
L_{n-1}^{m} (t)
=0 
\end{eqnarray}
for $n \ge 1$.

\subsection{Ladder operators for radial quantum number}
For obtaining ladder operators, we calculate the derivative of the generating function by $t$ as 
\begin{eqnarray}
\sum_{n=0}^{\infty}
\frac{d}{dt}
L_{n}^{m} (t)
\tau^n
=
-\frac{\tau}{1-\tau}
\sum_{n=0}^{\infty}
L_{n}^{m} (t)
\tau^n,
\end{eqnarray}
which becomes
\begin{eqnarray}
\sum_{n=0}^{\infty}
\frac{d}{dt}
L_{n}^{m} (t)
\tau^n
-
\sum_{n=0}^{\infty}
\frac{d}{dt}
L_{n}^{m} (t)
\tau^{n+1}
=
-
\sum_{n=0}^{\infty}
L_{n}^{m} (t)
\tau^{n+1}. \nonumber \\
\end{eqnarray}
Then, we obtain the identity for lowering $n$
\begin{eqnarray}
\frac{d}{dt}
L_{n}^{m} (t)
=
\left[
\frac{d}{dt}
-1
\right]
L_{n-1}^{m} (t).
\end{eqnarray}
However, this expression is not perfect, since the derivative operator remained in the right-hand side, which will be removed later.

Next, we construct the raising operator by calculating the derivative of the recurrence equation by $t$ as
\begin{eqnarray}
&& (n+1)
\left[
\frac{d}{dt}
-1
\right]
L_{n}^{m} (t)
+
L_{n}^{m} (t) \nonumber \\
&&
-
(2 n +m+1-t) 
\frac{d}{dt}
L_{n}^{m} (t) \nonumber \\
&&+
(n+m)
\frac{d}{dt}
L_{n-1}^{m} (t)
=0,
\end{eqnarray}
which becomes
\begin{eqnarray}
\left[
(n+m-t)
\frac{d}{dt}
+n
\right]
L_{n}^{m} (t)
=
\left[
(n+m)
\frac{d}{dt}
\right]
L_{n-1}^{m} (t). \nonumber \\
\end{eqnarray}
By using the lowering identity, this becomes
\begin{eqnarray}
&&(n+m-t)
\left[
\frac{d}{dt}
-1
\right]
L_{n-1}^{m} (t)
+n
L_{n}^{m} (t) \nonumber \\
&&=
(n+m)
\frac{d}{dt}
L_{n-1}^{m} (t), 
\end{eqnarray}
which gives the raising operator
\begin{eqnarray}
\left[
t\frac{d}{dt}
-t
+n+m+1
\right]
L_{n}^{m} (t)
=
(n+1)
L_{n+1}^{m} (t) \nonumber \\
\end{eqnarray}
for $n \ge 1$.
This expression is preferable, since the derivative operation appeared only in the left-side.
Together with this raising operator and the recurrence formula, we can eliminate $L_{n+1}^{m} (t)$ to obtain the lowering operator
\begin{eqnarray}
\left[
t\frac{d}{dt}
-n
\right]
L_{n}^{m} (t)
=
-
(n+m)
L_{n-1}^{m} (t),
\end{eqnarray}
while keeping $m$ unchanged.
These ladder operations for the associated Laguerre functions are consistent with those for Laguerre function in the limit of $m=0$.

\subsection{Ladder operators for orbital angular momentum}
The above formulas for raising and lowering the radial quantum numbers are known in literatures \cite{Arfken05,Whittaker62,Bateman53}, while we could not find appropriate formulas for raising and lowering quantum number $m$ for orbital angular momentum without affecting the radial quantum number of $n$.
Here, we derived these by direct calculations.
First, we obtain the raising operator by calculating
\begin{eqnarray}
L_n^{m+1}(t)
&=&
(-1)^{m+1}
\frac{d^{m+1}}{dt^{m+1}} 
L_{n+m+1}(t)
\\
&=&
-
\frac{d}{dt}
(-1)^{m}
\frac{d^{m}}{dt^{m}} 
L_{(n+1)+m}(t)
\\
&=&
-
\frac{d}{dt}
L_{n+1}^{m}(t)
\\
&=&
-
\left[
\frac{d}{dt}
-1
\right]
L_{n}^{m} (t).
\end{eqnarray}
Thus, the raising operator is described as
\begin{eqnarray}
\left[
\frac{d}{dt}
-1
\right]
L_{n}^{m} (t)
=
-L_n^{m+1}(t).
\end{eqnarray}

It was less straightforward to obtain the lowering operator.
As for preparations, we recognised several useful recurrence formulas
\begin{eqnarray}
L_n^{m+1}(t)
&=&
-
\left[
\frac{d}{dt}
-1
\right]
L_{n}^{m} (t), \\
L_n^{m+1}(t)
&=&
-
\frac{d}{dt}
L_{n+1}^{m}(t) , \\
L_n^{m+1}(t)
&=&
L_{n-1}^{m+1}(t)
+
L_n^{m}(t),  \\
L_n^{m}(t)
&=&
L_{n-1}^{m}(t)
+
L_n^{m-1}(t).
\end{eqnarray}

By using the recurrence formula, we obtain
\begin{eqnarray}
L_{n-1}^{m} (t)
=
\frac{2n+m+1-t}{n+m}
L_{n}^{m} (t)
-
\frac{n+1}{n+m}
L_{n+1}^{m} (t). \nonumber \\
\end{eqnarray}
Next, we use the raising operator for $n$ to obtain
\begin{eqnarray}
L_{n+1}^{m} (t)
=
\frac{1}{n+1}
\left[
t\frac{d}{dt}
-t
+n+m+1
\right]
L_{n}^{m} (t). \nonumber \\
\end{eqnarray}
By combining these equations, we obtain
\begin{eqnarray}
L_{n-1}^{m} (t)
=
\frac{1}{n+m}
\left[
n
-
t
\frac{d}{dt}
\right]
L_{n}^{m} (t).
\end{eqnarray}
Inserting this into Eq. (B33), we obtain the lowering operator
\begin{eqnarray}
\left[
t
\frac{d}{dt}
+m
\right]
L_{n}^{m} (t)
=
(n+m)
L_{n}^{m-1} (t).
\end{eqnarray}

\subsection{Orthogonality relationship}
The orthogonality relationship is obtained in a similar way by using the generating function
\begin{eqnarray}
\sum_{n=0}^{\infty}
L_{n}^{m} (t)
\tau^n
=
\frac{{\rm e}^{-\frac{t \tau}{1-\tau}} }
  {(1-\tau)^{m+1}},
\end{eqnarray}
and calculating the sum
\begin{eqnarray}
&&
\sum_{n=0}^{\infty}
\sum_{n'=0}^{\infty}
\tau^n
\tau'^{n'}
\int_{0}^{\infty}
dt
{\rm e}^{-t}
t^m
L_{n}^m(t)
L_{n'}^m(t)
\nonumber \\
&&=
\frac{1}{(1-\tau)^{m+1}(1-\tau')^{m+1}}
\int_{0}^{\infty}
dt
t^m
{\rm e}^{-t\frac{1-\tau\tau'}{(1-\tau)(1-\tau')}} \nonumber \\
&&=
\frac{m!}{(1-\tau \tau')^{m+1}}  \nonumber \\
&&=
\sum_{n=0}^{\infty}
\sum_{n'=0}^{\infty}
\frac{(n+m)!}{n!}
\delta_{n,n'}
\tau^n \tau'^{n'},
\end{eqnarray}
where we  used the binomial theorem
\begin{eqnarray}
\left(
\frac{1}{1-\tau \tau'}
\right)
^{m+1}
=
\sum_{n=0}^{\infty}
\frac{(n+m)!}{n! m!}
\tau^n \tau'^n.
\end{eqnarray}
Thus, we obtain the orthogonality relationship
\begin{eqnarray}
\int_{0}^{\infty}
dt
{\rm e}^{-t}
t^m
L_{n}^m(t)
L_{n'}^m(t)
=
\frac{(n+m)!}{n!}
\delta_{n,n'}.
\end{eqnarray}

\subsection{Rodrigues formula}
The Rodrigues formula for the associated Laguerre function is obtained by the direct calculations.
First, we calculate 
\begin{eqnarray}
L_n^m(t)&=&
(-1)^m
\frac{d^m}{dt^m} 
L_{n+m}(t) \nonumber \\
&=&
(-1)^m
\frac{d^m}{dt^m} 
\left[
\frac{1}{(n+m)!} 
{\rm e}^{t}
\frac{d^{n+m}}{dt^{n+m}}
\left(
t^{n+m}
{\rm e}^{-t}
\right)
\right]
\nonumber \\
&=&
\frac{d^m}{dt^m} 
\left[
\sum_{k=0}^{n+m}
\frac{1}{(n+m-k)!}
t^{n+m-k} \right. \nonumber \\
&& \ \ \left.
(-1)^{n-k}
\frac{(n+m)!}{(n+m-k)! k!}
\right]
\nonumber \\
&=&
\sum_{k=0}^{n}
\frac{1}{(n+m-k)!}
\frac{(n+m-k)!}{(n-k)!}
t^{n-k} \nonumber \\
&& \ \ (-1)^{n-k}
\frac{(n+m)!}{(n+m-k)! k!}
\nonumber \\
&=&
\sum_{k=0}^{n}
\frac{(n+m)!}{(n+m-k)! k! (n-k)!}
(-t)^{n-k}
\end{eqnarray}

On the other hand, we calculate 
\begin{eqnarray}
&& \frac{1}{n!}
t^{-m}
{\rm e}^{t}
\frac{d^n}{dt^n}
\left(
t^{n+m}
{\rm e}^{-t}
\right) \nonumber \\
&=&
\sum_{k=0}^{n}
\frac{(n+m)!}{k!(n-k)!(n+m-k)!}
(-t)^{n-k} \nonumber \\
\end{eqnarray}

By comparison, we obtain the Rodrigues formula
\begin{eqnarray}
L_n^m(t)
&=
\frac{1}{n!}
t^{-m}
{\rm e}^{t}
\frac{d^n}{dt^n}
\left(
t^{n+m}
{\rm e}^{-t}
\right).
\end{eqnarray}

\subsection{Integration formulas}
We also obtained integration formulas for the associated Laguerre functions:
\begin{eqnarray}
\int_{0}^{\infty}
da
{\rm e}^{-a}
a^m
L_{n}^m(a)
L_{n'}^m(a)
=
\frac{(n+m)!}{n!}
\delta_{n,n'},
\end{eqnarray}

\begin{eqnarray}
\int_{0}^{\infty}
da
{\rm e}^{-a}
a^m
L_{n}^m(a)
L_{n}^{m-1}(a)
=
\frac{(n+m)!}{n!},
\end{eqnarray}

\begin{eqnarray}
\int_{0}^{\infty}
da
{\rm e}^{-a}
a^{m+1}
L_{n}^{m+1}(a)
L_{n}^{m}(a)
=
\frac{(n+m+1)!}{n!}, \nonumber \\
\end{eqnarray}

\begin{eqnarray}
\int_{0}^{\infty}
da
{\rm e}^{-a}
a^{m+1}
L_{n}^m(a)
L_{n}^m(a)
=
\frac{(n+m)!}{n!}
(2n+m+1). \nonumber \\
\end{eqnarray}
These are useful to calculate the matrix elements.
We also obtained an identity, 
\begin{eqnarray}
&&
m
\int_{0}^{\infty}
da
{\rm e}^{-a}
a^{m-1}
L_{n}^m(a)
L_{n}^m(a)
+
\frac{(n+m)!}{n!}
\nonumber \\
&&
= 2
\int_{0}^{\infty}
da
{\rm e}^{-a}
a^m
L_{n}^m(a)
L_{n}^{m+1}(a).
\end{eqnarray}

\bibliography{QOAM}

\end{document}